\documentclass[11pt,preprint,showpacs,preprintnumbers,amsmath,amssymb]{revtex4}

\usepackage{graphicx}
\usepackage{dcolumn}
\usepackage{bm}
\usepackage{multirow}
\usepackage{graphicx}
\usepackage{float}
\usepackage{CJK}
\usepackage{caption}

\begin{document}

\title{X(3872) as a molecular $D\bar{D}^*$ state in the Bethe-Salpeter equation approach}

\author{Zhen-Yang Wang \footnote{e-mail: wangz-y@mail.bnu.edu.cn}}
\affiliation{\small{College of Nuclear Science and Technology, Beijing Normal University, Beijing 100875, China}}

\author{Jing-Juan Qi \footnote{e-mail: qijj@mail.bnu.edu.cn}}
\affiliation{\small{College of Nuclear Science and Technology, Beijing Normal University, Beijing 100875, China}}

\author{Chao Wang \footnote{e-mail: chaowang@nwpu.edu.cn}}
\affiliation{\small{Center for Ecological and Environmental Sciences, Key Laboratory for Space Bioscience \& Biotechnology, Northwestern Polytechnical University, Xi'an 710072, China}}

\author{Xin-Heng Guo \footnote{Corresponding author, e-mail: xhguo@bnu.edu.cn}}
\affiliation{\small{College of Nuclear Science and Technology, Beijing Normal University, Beijing 100875, China}}

\date{\today}

\begin{abstract}
 We discuss the possibility that the X(3872) can be a $D\bar{D}^*$ molecular bound state in the Bethe-Salpeter equation approach in the ladder and instantaneous approximations. We show that the $D\bar{D}^*$ bound state with quantum numbers $J^{PC}=1^{++}$ exists. We also calculate the decay width of $X(3872) \rightarrow \gamma J/\psi$ channel and compare our result with those from previous calculations.

\end{abstract}


\maketitle

\section{introduction}
  The $X(3872)$ was first observed by Belle collaboration in 2003 \cite{Choi:2003ue}, and later confirmed by CDF \cite{Acosta:2003zx}, D0 \cite{Abazov:2004kp}, and $BABAR$ collaborations \cite{Aubert:2004ns}. The new results of Belle collaboration show that $m_{X(3872)} = 3871.85 \pm 0.27(stat) \pm 0.19(syst)$ MeV and the upper limit on the width of $X(3872)$ is $\Gamma_{X(3872)} < 1.2$ MeV \cite{Choi:2011fc}. So far, several decay modes of the $X(3872)$ into $J/\psi \pi^+ \pi^-$, $J/\psi \pi^+ \pi^- \pi^0$, $\gamma J\psi$ and $\gamma \psi'$ have been identified \cite{Olive:2016xmw}, which give some constraints on the quantum numbers of this state. In particular, the decay mode $X(3872) \rightarrow \gamma J/\psi$ suggests the positive charge parity $C$ = + of this resonance. The quantum numbers of the $X(3872)$ are determined to be $J^P = 1^{++}$ based on angular correlations in $X(3872) \rightarrow J/\psi \pi^+ \pi^-$ by LHCb collaboration \cite{Aaij:2013zoa}.

  In the theory aspect, the nature of the $X(3872)$ is still a puzzle, and many theoretical models were proposed to explain the $X(3872)$ state. The $X(3872)$ is analyzed as a $D^0 \bar{D}^{*0}/\bar{D}^0 D^{*0}$ bound state \cite{Wang:2013kva,Thomas:2008ja,Braaten:2010mg,Baru:2013rta,Baru:2015nea,Close:2003sg,Voloshin:2003nt,Swanson:2003tb,Liu:2008fh,Liu:2008tn}, a tetraquark state \cite{Maiani:2004vq,Hogaasen:2005jv,Ebert:2005nc,Barnea:2006sd,Matheus:2006xi,Chiu:2006hd}, a hybrid charmonium ($c\bar{c}g$ gluonic hadrons) \cite{Close:2003mb,Li:2004sta}, and a charmonium ($c\bar{c}$) \cite{Barnes:2003vb,Eichten:2004uh,Kong:2006ni}, it has also been considered as a mixture of a charmonium with a $D^0 \bar{D}^{*0}/\bar{D}^0 D^{*0}$ component \cite{Meng:2005er,Suzuki:2005ha,Meng:2014ota}. Among the above models, the molecular state provides a plausible explanation since the $X(3872)$ can be identified as a weakly bound hadronic molecule which constituents are $D$ and $D^*$. The reason for this natural interpretation is that the mass of $X(3872)$ is very close to the $D^0 \bar{D}^{*0}$ threshold and hence is in analogy to the deuteron--a weakly bound state of the proton and the neutron.

  The radiative decay of the $X(3872)$ into $\gamma J/\psi$ is sensitive to its internal structure \cite{Swanson:2004pp,Barnes:2003vb}, and this decay channel has been studied in lots of literatures. The first observation of the $X(3872) \rightarrow \gamma J/\psi$ decay mode was reported by Belle collaboration \cite{Abe:2005ix}. Later on, this decay mode was confirmed by the $BABAR$ collaboration \cite{Aubert:2006aj} and again observed by Belle collaboration \cite{Bhardwaj:2011dj}. This decay mode received some early attention and was studied in Refs. \cite{Barnes:2003vb,Swanson:2004pp,Braaten:2003he,Braaten:2005ai,Aceti:2012cb}, assuming a charmonium state, a molecular state, or a mixture of a molecular state with a charmonium state.

  The Bethe-Salpeter (BS) equation is a formally exact equation to describe the relativistic bound state \cite{Salpeter:1951sz,Itzykson:1980,lurie-book} and has been applied to many theoretical studies concerning heavy mesons and heavy baryons \cite{Jin:1992mw,Guo:1996jj,Guo:1998ef,Guo:1998at,Guo:1999ss,Guo:2001wi,Guo:2007mm,Xie:2010zza,Feng:2012zze}. In this paper, we will work in the BS equation approach which can automatically include relativistic corrections comparing with the potential model which was applied in Ref. \cite{Zhang:2006ix} to investigate the possible states of $K\bar{K}$, $DK$, and $B\bar{K}$ in the framework of the nonrelativistic Schr$\ddot{o}$dinger equation with the potential between pseudoscalar mesons being derived from the relevant Lagrangian. We will try to investigate the possibility of $X(3872)$ as the $D\bar{D}^*$ molecular state with quantum numbers $J^{PC} = 1^{++}$. We will also study the decay of $X(3872)$ to $\gamma J/\psi$ in this picture.

  The paper is organized as follows. In Sec. \ref{sect-BS}, we establish the BS equation for the bound state of a vector meson and a pseudoscalar one. Then we discuss the interaction kernel. In Sec. \ref{sect-bound-state-of-DD*}, we discuss the normalization condition of the BS wave function and obtain the numerical results of the BS wave function. In Sec. \ref{sect-decay-width-of-DD*}, the decay of the $D\bar{D}^*$ bound state to $\gamma J/\psi$ final state is discussed and we give numerical results. Finally, Sec. \ref{sum-con} is devoted to summary and conclusion.

\section{the bethe-salpeter formalism for $D\bar{D}^*$ system}
\label{sect-BS}

In this section, we will review the general formalism of the BS equation and derive the BS equation for the system $D$ and $D^*$ mesons. We will also derive the normalization condition for the BS wave function. Let us start by defining the BS wave function for the bound state $|P\rangle$ of a vector and a pseudoscalar mesons as the following:
\begin{equation}\label{BS}
  \chi_P^\alpha(x_1,x_2,P) = \langle0|T\mathcal{D}^{*\alpha}(x_1)\mathcal{D}(x_2)|P\rangle = e^{-iPX}\chi_P^\alpha(x),
\end{equation}
where $\mathcal{D}^{*\alpha}(x_1)$ and $\mathcal{D}(x_2)$ are the field operators of the vector meson $D^*$ and the pseudoscalar meson $D$ at space coordinates $x_1$ and $x_2$, respectively, $P$ denotes the total momentum of the bound state with mass $M$ and velocity $v$, and the relative coordinate $x$ and the center-of-mass coordinate $X$ are defined by
\begin{equation}\label{co}
  X = \eta_1x_1 + \eta_2x_2, \quad x = x_1 - x_2,
\end{equation}
or inversely,
\begin{equation}
  x_1 = X + \eta_2x,   \quad   x_2 = X - \eta_1x,
\end{equation}
where $\eta_i = m_i/(m_1 + m_2)$, $m_i(i = 1, 2)$ is the mass of the $i$-th constituent particle. In momentum space, the BS wave function can be defined as
\begin{equation}\label{momentum BS function}
 \chi_P^\alpha(x_1,x_2,P) = e^{-iPX}\int\frac{d^4p}{(2\pi)^4}e^{-ipx}\chi_P^\alpha(p),
\end{equation}
where $p$ represents the relative momentum of the two constituents and $p= \eta_2 p_1-\eta_1 p_2$ (or $p_1=\eta_1P+p$,\quad $p_2=\eta_2P-p$).

The BS equation for the bound state of $X(3872)$ can be written in the following form:
\begin{equation}\label{BS equation}
  \chi_{P}^\alpha(p)=S^{\alpha\lambda}(p_1)\int\frac{d^4q}{(2\pi)^4}K_{\lambda\tau}(P,p,q)\chi_{P}^\tau(q)S(p_2),
\end{equation}
where $S^{\alpha\lambda}(p_1)$ and $S(p_2)$ are the propagators of $D^*$ and $D$, respectively, and $K_{\lambda\tau}(P,p,q)$ is the kernel which contains two-particle-irreducible diagrams. For convenience, we define $p_l (=p\cdot v)$ and $p_t^\mu(=p^\mu- p_lv^\mu)$ to be the longitudinal and transverse  projections of the relative momentum ($p$) along the bound state momentum ($P$). Then the $D^*$ propagator has the form
\begin{equation}\label{vector propagator}
  S^{\alpha\beta}(p_1)=\frac{-i\left[g^{\alpha\beta}-(\eta_1Mv+p_lv+p_t)^\alpha(\eta_1Mv+p_lv+p_t)^\beta/m_1^2\right]}{(\eta_1M+p_l)^2-w_1^2+i\epsilon},
\end{equation}
and the propagator of $D$ meson has the form
\begin{equation}\label{pseudoscalar propagator}
  S(p_2)=\frac{i}{(\eta_2M-p_l)^2-w_2^2+i\epsilon},
\end{equation}
where $\omega_{1(2)}=\sqrt{m_{1(2)}^2-p_t^2}$.

In general, for $D\bar{D}^*$ system, $\chi^\alpha_{P}(p)$ can be written as
\begin{equation}
\begin{split}
\chi^\alpha_{P}(p)=& f_1(p) \varepsilon^{\alpha\beta\mu\nu} g_{\mu\nu} \epsilon_\beta(P) + f_2(p)\varepsilon^{\alpha\beta\mu\nu} P_\mu P_\nu \epsilon_\beta(P) + f_3(p)\varepsilon^{\alpha\beta\mu\nu} p_\mu P_\nu \epsilon_\beta(P) + f_4(p)\varepsilon^{\alpha\beta\mu\nu} p_\mu p_\nu \epsilon_\beta(P),
\end{split}
\end{equation}
where $\epsilon_\beta(P)$ represents the polarization vector of the bound state and $f_i$($i=1,2,3,4$) are Lorentz-scalar functions. With the constraints imposed by parity and Lorentz transformations, it is easily to prove that $\chi^\alpha_{P}(p)$ can be simplified as
\begin{equation}
  \chi_P^\alpha(p) = \phi_P(p)\epsilon^\alpha,
\end{equation}
where the function $\phi_P(p)$ contains all the dynamics and is a Lorentz-scalar function of $p$.


As discussed in the introduction, we will study the $X(3872)$ as an $S$-wave bound state of the $D\bar{D}^*$ system. We use the field doublets $(D^{*0},\bar{D}^{*0})$, $(D^{*+},D^{*-})$, $(D^0,\bar{D}^0)$, and $(D^+,D^-)$, which correspond to the following expansions:
\begin{equation}\label{field-expand}
\begin{split}
D_1^* &= \int \frac{d^3p}{(2\pi)^3\sqrt{2E_{D^{*0}}}}(a_{D^{*0}}e^{-ipx}+a_{\bar{D}^{*0}}^\dagger e^{ipx}),\\
D_2^* &= \int \frac{d^3p}{(2\pi)^3\sqrt{2E_{D^{*\pm}}}}(a_{D^{*+}}e^{-ipx}+a_{D^{*-}}^\dagger e^{ipx}),\\
D_3 &= \int\frac{d^3p}{(2\pi)^3\sqrt{2E_{D^0}}}(a_{D^0}e^{-ipx}+a_{\bar{D}^0}^\dagger e^{ipx}),\\
D_4 &= \int\frac{d^3p}{(2\pi)^3\sqrt{2E_{D^\pm}}}(a_{D^+}e^{-ipx}+a_{D^-}^\dagger e^{ipx}).
\end{split}
\end{equation}

Since the isospin quantum number of $X(3872)$ is zero, if we assume that it is composed of $D\bar{D}^{*}$, the flavor wave function of the $X(3872)$ can be represented as in Refs. \cite{Y.R.Liu:2008,X.Liu:2009}

\begin{equation}
|X(3872)\rangle = \frac{1}{\sqrt{2}}[|\bar{D}^{*0}D^0\rangle - |\bar{D}^{0}D^{*0}\rangle].
\end{equation}

Let us now project the bound states on the field operators $D_1^*$, $D_2^*$, $D_3$ and $D_4$. From Eq. (\ref{field-expand}) we have
\begin{equation}\label{projection-of-BS-state}
\begin{split}
  &\langle 0| {\rm T}\,\{ D^{*\dag}_i(x_1) D_j(x_2) \} |P\rangle_{I,I_3} = C_{(I,I_3)}^{ij}\chi_P^{(I)}(x_1,x_2),\\
  &\langle 0| {\rm T}\,\{ D^{\dag}_i(x_1) D^*_j(x_2) \} |P\rangle_{I,I_3} = C_{(I,I_3)}^{ij}\chi_P^{(I)}(x_1,x_2),
\end{split}
\end{equation}
where $C_{(I,I_3)}^{ij}$ ($i$, $j$ =1, 2, 3, 4) is the isospin coefficient. The coefficients $C_{(I,I_3)}^{ij}$ for the isoscalar state are
\begin{equation}\label{isoscalar-coefficient-0}
C_{(0,0)}^{13}= 1/\sqrt{2}, \quad C_{(0,0)}^{31}= -1/\sqrt{2}.
\end{equation}

\begin{figure}[ht]
\centering
    \rotatebox{0}{\includegraphics*[width=0.30\textwidth]{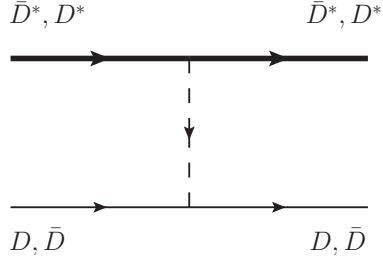}}
    \caption{The direct-channel Feynman diagram for the $D\bar{D}^{*}$ system at the tree level. The thick line represents the vector state $D^*$ and $\bar{D}^{*}$ while the thin line stands for $D$ and $\bar{D}$.}
  \label{direct-channel-fig}
\end{figure}

\begin{figure}[ht]
\centering
    \rotatebox{0}{\includegraphics*[width=0.35\textwidth]{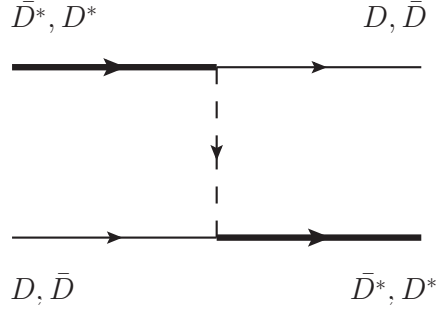}}
    \caption{The cross-channel Feynman diagram for the $D\bar{D}^{*}$ system at the tree level. Notations are the same as in Fig. \ref{direct-channel-fig}.}
  \label{cross-channel-fig}
\end{figure}

The kernel of the BS equation for the $X(3872)$ can be derived from the meson exchange Feynman diagrams for the $D\bar{D}^{*}$ system at the tree level which are shown in Fig. \ref{direct-channel-fig} and Fig. \ref{cross-channel-fig}.



Based on the chiral symmetry~\cite{Wise:1992hn,Isola:2003fh}, the Lagrangians for the interactions among $D(D^\ast)$ mesons and light pseudoscalar, scalar or vector
mesons are
\begin{equation}
  \begin{split}
   \mathcal{L}_{\mathcal{D}\mathcal{D}^*\mathbb{P}} &= -ig_{DD^*\mathbb{P}}({\mathcal{D}}_a{\mathcal{D}}_{\mu b}^{* \dagger} -{\mathcal{D}}_{\mu a}^*{\mathcal{D}}_b^\dagger)\partial^\mu\mathbb{P}_{ab}, \\
\mathcal{L}_{{\mathcal{D}}{\mathcal{D}}^*\mathbb{V}} &= -2f_{{D}{D}^*\mathbb{V}} \varepsilon_{\mu\nu\alpha\beta}(\partial^\mu\mathbb{V}^\nu)_{ab} [({\mathcal{D}}_a^\dagger\partial^\alpha{\mathcal{D}}_b^{*\beta} -\partial^\alpha{\mathcal{D}}_a^\dagger{\mathcal{D}}_b^{*\beta}) -({\mathcal{D}}_a^{*\beta\dagger}\partial^\alpha{\mathcal{D}}_b -\partial^\alpha{\mathcal{D}}_a^{*\beta\dagger}{\mathcal{D}}_b)],\\
   \mathcal{L}_{{\mathcal{DD}}\sigma} &= -2m_{D} g_\sigma{\mathcal{D}}_a {\mathcal{D}}_a ^\dagger \sigma, \quad {\mathcal{L}}_{{\mathcal{D}}^*{\mathcal{D}}^*\sigma} = 2m_{{D}^*}
g_\sigma{\mathcal{D}}_a^{*\alpha}{\mathcal{D}}_{\alpha a}^{*\dagger}\sigma,\\
\mathcal{L}_{{\mathcal{DD}}\mathbb{V}} &= -ig_{{DD}\mathbb{V}}({\mathcal{D}}_a^\dagger\partial_\mu {\mathcal{D}}_b-{\mathcal{D}}_b \partial_\mu{\mathcal{D}}_a^\dagger)(\mathbb{V}^\mu)_{ab},\\
   \mathcal{L}_{\mathcal{D}^*\mathcal{D}^*\mathbb{V}} &= ig_{D^*D^*\mathbb{V}}({\mathcal{D}}_a^{*\nu\dagger}\partial^\mu{\mathcal{D}}_{\nu,b}^*-{\mathcal{D}}_{\nu,b}^*\partial^\mu{\mathcal{D}}_a^{*\nu\dagger})
   (\mathbb{V}_\mu)_{ab}+4if_{D^*D^*\mathbb{V}}\mathcal{D}_{\mu, a}^{*\dagger}{\mathcal{D}}_{\nu, b}^*(\partial^\mu\mathbb{V}^\nu-\partial^\nu \mathbb{V}^\mu)_{ab},
  \end{split}
\end{equation}
where $a$, $b$ denote the light quark flavour indices, The octet pseudoscalar $\mathbb{P}$ and the nonet vector $\mathbb{V}$ meson matrices are defined as
\begin{equation}
\label{eq:pseudo}
\mathbb{P}=\left(
\begin{array}{ccc}\frac{\pi^0}{\sqrt{2}}+\frac{\eta}{\sqrt{6}}&\pi^+&K^+\\
\pi^-&-\frac{\pi^0}{\sqrt{2}}+\frac{\eta}{\sqrt{6}}&K^0\\
K^-&\bar{K}^0&-\frac{2\eta}{\sqrt{6}}\\
\end{array} \right),
\end{equation}
and
\begin{equation}\label{eq:vector}
\mathbb{V}=\left( \begin{array}{ccc}\frac{\rho^0}{\sqrt{2}}+\frac{\omega}{\sqrt{2}}&\rho^+&K^{*+}\\
\rho^-&-\frac{\rho^0}{\sqrt{2}}+\frac{\omega}{\sqrt{2}}&K^{*0}\\
K^{*-}&\bar{K}^{*0}&\phi\\
\end{array} \right),
\end{equation}
respectively, and the coupling constants are given as
\begin{equation}\label{constants}
\begin{split}
g_{{ D}{ D}^*\mathbb{P}} &= \frac{2g}{f_\pi}\sqrt{m_{ D}m_{ D^*}}, \quad g_{{ D}{ D}\mathbb{V}}=g_{{ D}^*{ D}^*\mathbb{V}}=\frac{\beta g_{\mathbb{V}}}{\sqrt{2}},\\
f_{{ D}{ D}^*\mathbb{V}} &= \frac{f_{{ D}^*{ D}^*\mathbb{V}}}{m_{ D^*}} =\frac{\lambda g_{\mathbb{V}}}{\sqrt{2}}, \quad g_{\mathbb{V}}=\frac{m_\rho}{f_\pi},\\
g_\sigma &=\frac{g_\pi}{2\sqrt{6}},\quad g=0.59,\quad \beta=0.9,\\
\lambda &= 0.56 \ {\rm GeV}^{-1},\quad f_\pi=132 \ {\rm MeV},\quad g_\pi=3.73.
\end{split}
\end{equation}

From the above observations, at the tree level, in the $t$-channel we have the following kernel for the BS equation in the so-called lader approximation (see Figs. \ref{direct-channel-fig} and \ref{cross-channel-fig} for direct and crossed channels, respectively):
\begin{equation}\label{kernel-directly-channel-Sigma}
  \bar{K}^{\lambda\tau}_{direct}(p_1,p_2;q_2,q_1;m_\sigma)=(2\pi)^4\delta^4(q_1+q_2-p_1-p_2)4g_{\sigma}^2 m_{D^*}m_{D} \Delta(k,m_\sigma) g^{\lambda\tau},
\end{equation}
\begin{equation}\label{kernel-directly-channel-vector}
  \bar{K}^{\lambda\tau}_{crossed}(p_1,p_2;q_2,q_1;m_P)=-(2\pi)^4\delta^4(q_1+q_2-p_1-p_2)g_{DD^*\mathbb{P}}^2k^\lambda k^\tau \Delta(k,m_P),
\end{equation}
\begin{equation}\label{kernel-directly-channel-vectors}
\begin{split}
  \bar{K}^{\lambda\tau}_{direct}(p_1,p_2;q_2,q_1;m_V)=&(2\pi)^4\delta^4(q_1+q_2-p_1-p_2)\bigg\{ g_{D^* D^* \mathbb{V}}g_{{DD}\mathbb{V}} (p_1 + q_1)_\mu  (p_2 + q_2)_\nu  \Delta^{\mu\nu}(k,m_V) g^{\lambda\tau}\\
  &+4f_{D^*D^* \mathbb{V}}g_{{DD}\mathbb{V}} \Big[k^\lambda (p_2 + q_2)_\mu \Delta^{\tau\mu}(k,m_V)-k^{\tau} (p_2 + q_2)_\nu \Delta^{\lambda\nu}(k,m_V) \Big]\bigg\},
\end{split}
\end{equation}
\begin{equation}\label{kernel-directly-channel-pseudoscalar}
\begin{split}
  \bar{K}^{\lambda\tau}_{crossed}(p_1,p_2;q_2,q_1;m_V)=&(2\pi)^4\delta^4(q_1+q_2-p_1-p_2)4 f_{{D}{D}^*\mathbb{V}}^2 \epsilon^{\mu\beta\sigma\lambda}\epsilon^{\nu\rho\gamma\tau}k_\mu k_\nu (p_1 + q_2)_\sigma(q_1 + p_2)_\gamma \Delta_{\beta\rho}(k,m_V), \\
\end{split}
\end{equation}
where $m_\sigma$, $m_P$ and $m_V$ represent the masses of the exchanged $\sigma$, pseudoscalar light meson and vector light meson, respectively. $\Delta^{\mu\nu}$ represents the propagator for a vector meson and $\Delta$ represents pseudoscalar or scalar meson propagator, and they have the following forms:
\begin{equation}
\begin{split}
  \Delta^{\mu\nu} &= \frac{-i}{k^2 - m_V^2}\left(g_{\mu\nu} - \frac{k_\mu k_\nu}{m_V^2} \right),\\
  \Delta &= \frac{i}{k^2 - m_{\sigma(P)}^2}.\\
\end{split}
\end{equation}

In order to describe the phenomena in the real world, we should include a form factor at each interacting vertex of hadrons to include the finite-size effects of these hadrons. For the meson-exchange case, the form factor is assumed to take the following form:
\begin{equation}\label{form factor}
  F(k)=\frac{\Lambda^2-m^2}{\Lambda^2-k^2},
\end{equation}
where $\Lambda$, $m$ and $k$ represent the cutoff parameter, mass of the exchanged meson and momentum of the exchanged meson, respectively.

From Eqs. (\ref{BS equation}-\ref{pseudoscalar propagator}) and Eqs. (\ref{kernel-directly-channel-Sigma}-\ref{kernel-directly-channel-pseudoscalar}), we have
\begin{equation}\label{full BS equation}
  \begin{split}
  \phi_P(p)& = \frac{-i}{\left[(\eta_1M+p_l)^2-\omega_1^2\right]\left[(\eta_2M-p_l)^2-\omega_2^2\right]p_t\cdot p_t}\int\frac{d^4q}{(2\pi)^4} \\
  &\times\Bigg\{\frac{g_{DD^*\mathbb{P}}^2 (k\cdot p_t)^2}{3(k^2-m_\eta^2)}F_{m_\eta}^2(k)-\frac{g_{DD^*\mathbb{P}}^2k\cdot p_1 k\cdot p_t p_1\cdot p_t}{3m_1^2(k^2-m_\eta^2)}F_{m_\eta}^2(k)+\frac{g_{DD^*\mathbb{P}}^2(k\cdot p_t)^2}{k^2-m_\pi^2}F_{m_\pi}^2(k) \\
  &  +\frac{4g_\sigma^2 m_2(p_1\cdot p_t)^2}{m_1(k^2-m_\sigma^2)}F_{m_\sigma}^2(k)-\frac{4g_\sigma^2 m_1 m_2 p_t \cdot p_t}{(k^2-m_\sigma^2)}F_{m_\sigma}^2(k)+\frac{g_{DD\mathbb{V}}g_{DD^*\mathbb{V}}k\cdot(p_1+q_1)k\cdot(p_2+q_2)(p_1\cdot p_t)^2}{m_1^2m_\omega^2(k^2-m_\omega^2)}F_{m_\omega}^2(k)\\
  &+\frac{g_{DD\mathbb{V}}g_{D^*D^*\mathbb{V}}k\cdot(p_1+q_1)k\cdot(p_2+q_2)(p_1\cdot p_t)^2}{m_1^2m_\rho^2(k^2-m_\rho^2)}F_{m_\rho}^2(k)-\frac{g_{DD^*\mathbb{P}}^2k\cdot p_1 k\cdot p_t p_1\cdot p_t}{m_1^2(k^2-m_\pi^2)}F_{m_\pi}^2(k)\\
  &+\frac{g_{DD\mathbb{V}}f_{D^*D^*\mathbb{V}}k\cdot p_t p_1\cdot(p_2+q_2)p_1\cdot p_t}{m_1^2(k^2-m_\omega^2)}F_{m_\omega}^2(k)+\frac{g_{DD\mathbb{V}}f_{D^*D^*\mathbb{V}}k\cdot p_t P_1\cdot(p_2+q_2)p_1\cdot p_t}{m_1^2(k^2-m_\rho^2)}F_{m_\rho}^2(k)\\
  &-\frac{g_{DD\mathbb{V}}g_{D^*D^*\mathbb{V}}(p_1\cdot p_t)^2(p_1+q_1)\cdot(p_2+q_2)}{m_1^2(k^2-m_\omega^2)}F_{m_\omega}^2(k)-\frac{g_{DD\mathbb{V}}g_{D^*D^*\mathbb{V}}(p_1\cdot p_t)^2(p_1+q_1)\cdot(p_2+q_2)}{m_1^2(k^2-m_\rho^2)}F_{m_\rho}^2(k)\\
  &-\frac{4g_{DD\mathbb{V}}f_{D^*D^*\mathbb{V}}k\cdot p_1 p_1\cdot p_t (p_2+q_2)\cdot p_t}{m_1^2(k^2-m_\omega^2)}F_{m_\omega}^2(k)-\frac{4g_{DD\mathbb{V}}f_{D^*D^*\mathbb{V}}k\cdot p_1 p_1\cdot p_t (p_2+q_2)\cdot p_t}{m_1^2(k^2-m_\rho^2)}F_{m_\rho}^2(k)\\
  &-\frac{g_{DD\mathbb{V}}g_{D^*D^*\mathbb{V}}k\cdot(p_1+q_1)k\cdot(p_2+q_2)p_t^2}
  {k^2-m_\omega^2}F_{m_\omega}^2(k)-\frac{g_{DD\mathbb{V}}g_{D^*D^*\mathbb{V}}k\cdot(p_1+q_1)k\cdot(p_2+q_2)p_t^2}
  {k^2-m_\rho^2}F_{m_\rho}^2(k)\\
  &+\frac{g_{DD\mathbb{V}}g_{D^*D^*\mathbb{V}}(p_1+q_1)\cdot(p_2+q_2)p_t\cdot p_t}
  {k^2-m_\omega^2}F_{m_\omega}^2(k)+\frac{g_{DD\mathbb{V}}g_{D^*D^*\mathbb{V}}(p_1+q_1)\cdot(p_2+q_2)p_t\cdot p_t}
  {k^2-m_\rho^2}F_{m_\rho}^2(k)\\
  &-\left[\frac{4f_{DD^*\mathbb{V}}^2(p_1\cdot p_t)}{k^2-m_\omega^2}F_{m_\omega}^2(k)+\frac{4f_{DD^*\mathbb{V}}^2(p_1\cdot p_t)}{k^2-m_\rho^2}F_{m_\rho}^2(k)\right]\times\Big[k\cdot(p_1+q_2)k\cdot(q_1+p_2)p_1\cdot p_t\\
  &-k^2(p_1+q_2)\cdot(q_1+p_2)p_1\cdot p_t-k\cdot(p_1+q_2)k\cdot p_t p_1\cdot(q_1+p_2)-k\cdot p_1k\cdot(q_1+p_2)(p_1+q_2)\cdot p_t\\
  &+k^2p_1\cdot(q_1+p_2)(p_1+p_2)\cdot p_t+k\cdot p_1k\cdot p_t(p_1+q_2)\cdot(q_1+p_2)\Big]\\
  &-\left[\frac{4f_{DD^*\mathbb{V}}^2}{k^2-m_\omega^2}F_{m_\omega}^2(k)+\frac{4f_{DD^*\mathbb{V}}^2}{k^2-m_\rho^2}
  F_{m_\rho}^2(k)\right]\times\Big[-(p_1+q_2)\cdot(q_1+p_2)(k\cdot p_t)^2\\
  &+k\cdot(q_1+p_2)(p_1+q_2)\cdot p_t k\cdot p_t-k\cdot(p_1+q_2)k\cdot(q_1+p_2)p_t\cdot p_t-k\cdot(p_1+q_2)k\cdot(q_1+p_2)p_t\cdot p_t\\
  &+k^2(p_1+q_2)\cdot(q_1+p_2)p_t\cdot p_t-k^2(p_1+q_2)\cdot p_t(q_1+p_2)\cdot p_t\Big]\Bigg\} \phi_P(q),
\end{split}
\end{equation}
where $F_{m_\sigma}(k)$, $F_{m_\eta}(k)$ and so on represent the form factors for different exchanged mesons.

Define $\tilde{\phi}_P(|p_t|) = \int \frac{dp_l}{2\pi}\phi_P(p_l,p_t)$, $\tilde{\phi}_P(|p_t|)$ depends only on the norm of $p_t$, $|p_t|$. Therefore, after completing the azimuthal
integration, the above BS equation becomes a one dimensional integral equation, which reads
\begin{equation}
  \tilde{\phi}_P(|p_t|) = \int d|p_t| \left(V_1(|p_t|,|q_t|)+V_2(|p_t|,|q_t|)\right)\tilde{\phi}_P(|q_t|),
\end{equation}
the expressions for $V_1(|p_t|,|q_t|)$ and $V_2(|p_t|,|q_t|)$ are given in Appendix \ref{appendix}.

\section{Solution of the BS equation for X(3872)}
\label{sect-bound-state-of-DD*}

In this part, we will solve the BS equation numerically. To find out the bound state of the $D \bar{D}^{*}$ system, one only needs to solve the homogeneous BS equation. However, when we want to calculate physical quantities such as the decay width we have to face the problem of the normalization of the BS wave function. In the following we will discuss the
normalization of the BS wave function $\chi_P(p)$. Following Ref. \cite{lurie-book} one can write down the normalization condition as
\begin{equation}\label{normalization equation}
 i\int\frac{d^4pd^4q}{(4\pi)^8}\bar{\chi}_\alpha(p) \frac{\partial}{\partial P^0} I^{\alpha\beta}_P(p,q)\chi_\beta(q) = 2P_0,
\end{equation}
where $P^0 = E$, $I_P^{\alpha\beta}(p,q) = (2\pi)^4\delta^4(p-q)\left(S^{\alpha\beta}(p_1,m_1)\right)^{-1}S^{-1}(p_2,m_2)$. $\left(S^{\alpha\beta}(p_1,m_1)\right)^{-1}$ and $S^{-1}(p_2,m_2)$ have the following form:
\begin{equation}\label{inverse of vector propagator}
  \left(S^{\alpha\beta}(p_1,m_1)\right)^{-1} = i\left[g^{\alpha\beta}\left(p_1^2-m_1^2\right)-p_1^\alpha p_1^\beta\right],
\end{equation}
\begin{equation}\label{inverse of pseudoscalar propagator}
  S^{-1}(p_2,m_2) = -i\left(p_2^2-m_2^2\right).
\end{equation}

Inserting Eqs. (\ref{inverse of vector propagator}) and (\ref{inverse of pseudoscalar propagator}) into Eq. (\ref{normalization equation}), the normalization condition can be written in the following form:
\begin{equation}\label{normalization-BS-function}
  \begin{split}
  i\int\frac{d^4p}{(2\pi)^4}\bar{\phi}_P(p)\Big\{& -6\eta_1(\eta_1M+pl)\left[(\eta_2M-pl)^2+p_t^2-m_2^2\right]\\
  &+\left[-3(\eta_1M+p_l)^2-2p_t^2+3m_1^2\right]2\eta_2(\eta_2M-p_l) \Big\}\phi_P(q) =2P_0.
  \end{split}
\end{equation}
Substituting Eq. (\ref{full BS equation}) into Eq. (\ref{normalization-BS-function}) and completing the azimuthal integration we have
\begin{equation}
\begin{split}
 &\int d|q_t|\tilde{\phi}_P(|q_t|)^2\Big\{V_1(|p_t|,|q_t|)\big[6\eta_1\omega_1\left(M^2+2M\omega_1+\omega_1^2-|p_t|^2-m_2^2\right)\\
 &+2\eta_2\left(M+\omega_1\right)\left(-3\omega_1^2+2|p_t|^2+3m_1^2\right)\big]+V_2(|p_t|,|q_t|)\big[-6\eta_1\left(M+\omega_2\right)\left(\omega_2^2-|p_t|^2-m_2^2\right)\\
 &-2\eta_2\omega_2\left(-3M^2-6M\omega_1-3\omega_1^2+2|p_t|^2+3m_1^2\right)\big]\Big\}=2M.
\end{split}
\end{equation}

It can be seen from Eq. (\ref{full BS equation}) that there is one parameter in our model, the cutoff $\Lambda$, which contains the information about the non-point interaction due to the structure of hadrons at the interaction vertices. Although the value of $\Lambda$ cannot be exactly determined and depends on the specific process, it should be typically the scale of low-energy physics, which is about 1 GeV. In Ref. \cite{Liu:2008fh}, Liu $et$ $al.$ claimed that the bound state is not present for values of $\Lambda <$ 5.8 GeV when taking into account both pion and sigma meson-exchange potentials, where the upper value is large compared to the typical hadronic scale of $\Lambda \simeq 1$ GeV. Later, Liu $et$ $al.$ \cite{Liu:2008tn} showed that when including the exchange forces from the $\pi$, $\eta$ $\sigma$, $\rho$ and $\omega$ mesons there exists a molecular state when $\Lambda\sim$0.55 GeV. In Ref. \cite{Zhao:2014gqa}, the authors found there exists a bound-state solution when the cutoff parameter changes from 1.1 to 1.3 GeV in the effective potential model. In this work, we shall treat the cutoff $\Lambda$ in the form factors as a parameter varying in a wider range (0.5-4.8) GeV, in which we try to search for possible solutions of the $D\bar{D}^*$ bound states.

\begin{figure}[ht]
\centering
    \rotatebox{0}{\includegraphics*[width=0.60\textwidth]{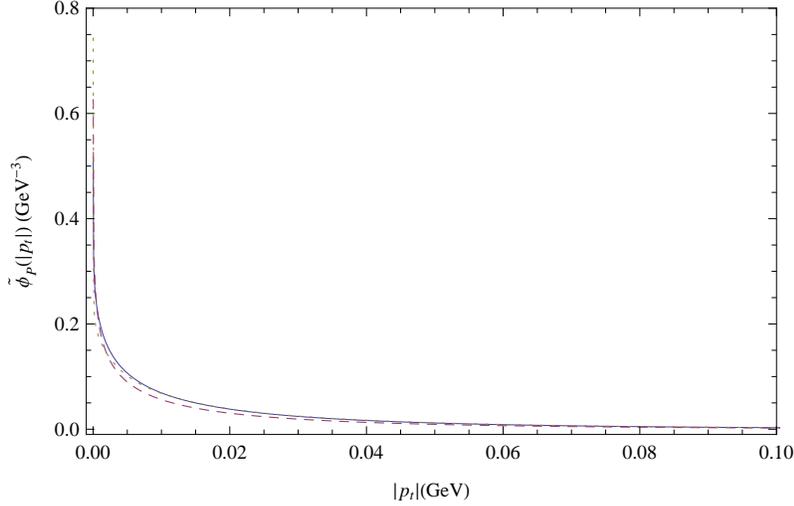}}
    \caption{Numerical result for the BS wave function $\tilde{\phi}_P(|p_t|)$ for the bound state of $D\bar{D}^*$. The solid, dashed and dotted lines correspond to $\Lambda$= 0.69 GeV, 0.83 GeV, and 0.96 GeV, respectively. The values of $\tilde{\phi}_P(|p_t|)$ when $|p_t|=0$ are 0.516, 0.625, and 0.742 $\mathrm{GeV}^{-3}$, respectively.}
  \label{numerical wave function}
\end{figure}

Let us first solve the BS bound state equation (\ref{full BS equation}) numerically. We discretize the integral equation (\ref{full BS equation}) (where we use the covariant instantaneous approximation, $p_l=q_l$) into a matrix eigenvalue equation by the Gaussian quadrature method. For each pair of trail values of the cutoff $\Lambda$ and the binding energy $E_b$ of the $D\bar{D}^*$ system (which is defined as $E_b = E - E_D - E_{D^*}$), we will obtain all the eigenvalues of this eigenvalue equation. The eigenvalue closest to 1.0 for a pair of $\Lambda$ and $E_b$ will be selected out and called ``the trial eigenvalue". In the study of the $D\bar{D}^*$ bound state, we choose to work in the rest frame of the bound state in which $P=(M,0)$, and we use $M=3872.2$ MeV, $M_{D^0}=1864.8$ MeV, $M_{D^{*0}}=2006.8$ MeV \cite{Patrignani:2016xqp}. We find the cutoff can be 0.69-0.96 GeV in our calculation. Because $E_b$ is very small compared with the masses of $D^0$ and $D^{*0}$ mesons, we find that the impact of different values of $E_b$ is particularly small and can be ignored. The numerical result for the BS wave function $\tilde{\phi}_P(|p_t|)$ for the bound state of $D\bar{D}^{*}$ is plotted in Fig. \ref{numerical wave function}, in which the solid, dashed and dotted lines correspond to $\Lambda$ = 0.69 GeV, 0.83 GeV, and 0.96 GeV, respectively. It can be seen from Fig. \ref{numerical wave function} that the numerical solutions of the BS wave function for different values of $\Lambda$ are very close to each other.

\section{The decay of X(3872) $\rightarrow$ $\gamma J/\psi$}
\label{sect-decay-width-of-DD*}

\begin{figure}
  \includegraphics[height = 4cm, width = 15 cm]{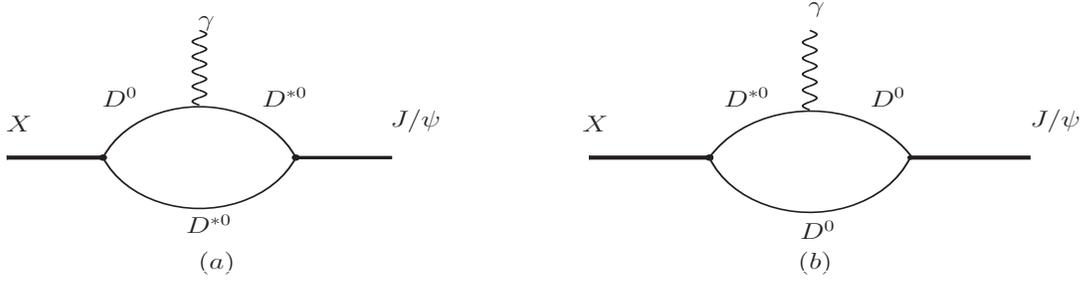}\\
  \caption{Diagrams contributing to the radiative transition $X(3872) \rightarrow \gamma J/\psi$.}
  \label{decay-feynman-diagarams}
\end{figure}

After obtaining the BS wave function, we can calculate some physical properties of the molecular bound state which can be measured in experiments. One of the most important properties is the decay width. The bound state of the X(3872) system can decay to $\gamma J/\psi$ via the Feynman diagrams in Fig. \ref{decay-feynman-diagarams}. In the following we will write down the decay amplitude and calculate the decay width using the solution of the one-dimensional BS equation obtained in the previous section. The effective Lagrangian for the radiative decay X(3872) $\rightarrow$ $\gamma J/\psi$ is \cite{Dong:2008gb} :
\begin{equation}
  \begin{split}
    &\mathcal{L}_{D^{*0}D^0\gamma}  = \frac{e}{4} g_{D^{*}D\gamma} \epsilon^{\mu\nu\alpha\beta} F_{\mu\nu}(x) \bar{\mathcal{D}}^{*0}_{\alpha\beta}(x) \mathcal{D}^0(x) + h.c.,\\
    &\mathcal{L}_{J_{\psi}D^0D^0}  = ig_{J_{\psi}DD} J_{\psi}^{\mu}(x) \left( \mathcal{D}^0(x)\partial_{\mu} \bar{\mathcal{D}}^0(x) - \bar{\mathcal{D}}^0(x) \partial_{\mu} \mathcal{D}^0(x) \right),\\
    &\mathcal{L}_{J_{\psi}D^{*0}D^{*0}}  = i g_{J_{\psi}D^*D^*} \left( J^{\mu\nu}_{\psi}(x) \bar{\mathcal{D}}^{*0}_{\mu}(x) \mathcal{D}^{*0}_{\nu}(x) + J^{\mu}_{\psi}(x) \bar{\mathcal{D}}^{*0\nu}(x)\mathcal{D}^{*0}_{\mu\nu}(x) + J_{\psi}^{\nu}(x)\bar{\mathcal{D}}^{*0}_{\mu\nu}(x)\mathcal{D}^{*0\mu}(x)\right),
  \end{split}
\end{equation}
where $F_{\mu\nu} = \partial_\mu A_\nu - \partial_\nu A_\mu$ and $M_{\mu\nu} = \partial_\mu M_\nu - \partial_\nu M_\mu$ is the stress tensor of the vector mesons with $M = \mathcal{D}^{*0}, J_{\psi}$ (in the Lagrangian we denote $J/\psi$ by $J_{\psi}$), $\epsilon^{\mu\nu\alpha\beta}$ is levi-civita symbol. In the present calculation we will use the following values of the coupling constants \cite{Dong:2008gb}:
\begin{equation}
  g_{J_{\psi}DD}=g_{J_{\psi}D^*D^*}=6.5, \quad g_{D^{*0}D^0\gamma}=2.
\end{equation}

The differential decay width of the bound state can be written as
\begin{equation}
  d\Gamma = \frac{1}{32\pi^2} |\mathcal{M}|^2 \frac{|\mathbf{q}|}{E^2} d\Omega,
\end{equation}
where $|\mathbf{q}|$ is the norm of the three-momentum of the particles in the final state in the rest of the initial bound state. $\mathcal{M}$ is the Lorentz-invariant decay amplitude of the process.

According to the above interactions, the decay X(3872) $\rightarrow$ $\gamma J/\psi$ induced by $D^*$ exchange is shown in Fig. 4(a). We can write down the amplitude as
\begin{equation}
\begin{split}
  \mathcal{M}_{D^*}&=-ie g_{D^*D\gamma}g_{J_{\psi}D^*D^*} \epsilon^{\kappa\lambda\alpha\beta}\int \frac{d^4p}{(4\pi)^4}F(|\mathbf{k}|)^2\Big[(-ik_\alpha)(iq_{1\kappa})(iq_{2}^\rho)\Delta_\beta^\gamma(k,D^*)-(-ik_\alpha)(iq_{1\kappa})(iq_{2}^\nu)\Delta_{\beta\nu}(k,D^*)g^{\rho\gamma}\\
  &+(-ik_\alpha)(ik^{\gamma})(iq_{1\kappa})\Delta_{\beta}^{\rho}(k,D^*)-(-ik_\alpha)(ik^{\rho})(iq_{1\kappa})\Delta_{\beta}^{\rho}(k,D^*)+
  (-ik_\alpha)(-ip_{2\mu})(iq_{1\kappa})\Delta_{\beta}^{\mu}(\kappa,D^*)g^{\rho\gamma}\\
  &-(-ik_\alpha)(-ip^{2\gamma})(iq_{1\kappa})\Delta_{\beta}^{\rho}(k,D^*)\Big]_{k=q-p}\epsilon_\rho(P)\epsilon_\lambda(q_1)\epsilon_\gamma(q_2)\phi_{P}(p),\\
  \end{split}
\end{equation}
where $q_1 (q_2)$ is the momentum of $\gamma (J_{\psi})$ and $q= \eta_2 q_1- \eta_1 q_2$ which is not the relative momentum of particles in the final state (note that $\eta_1$ and $\eta_2$ are defined as $\eta_i = m_i/(m_1 +m_2)$, and $m_1$ and $m_2$ are the masses of the component particles of the initial bound state but not the final states), $\epsilon_\rho(P)$, $\epsilon_\lambda(q_1)$ and $\epsilon_\gamma(q_2)$ are the polarization vectors of X(3872), $\gamma$ and $J_{\psi}$, respectively.

Similarly, the diagram for X(3872) $\rightarrow$ $\gamma J/\psi$ through exchanging the $D$ meson is shown in Fig. 4(b). One can write the amplitude as
\begin{equation}
  \begin{split}
  \mathcal{M}_D&=-ieg_{D^*D\gamma}g_{J_{\psi}DD}\epsilon^{\kappa\lambda\alpha\beta}\int \frac{d^4p}{(4\pi)^4}F(|\mathbf{k}|)^2\Big[(ik_{\mu})(-ip_{1\alpha})(iq_{1\kappa})-(-ip_{1\alpha})(-ip_{2\mu})(iq_{1k})\Big]_{k=q-p}\\
  &\times g_\beta^\rho g^{\gamma\mu}\Delta(k,D)\epsilon_\rho(P)\epsilon_\lambda(q_1)\epsilon_\gamma(q_2)\phi_{P}(p).\\
  \end{split}
\end{equation}



In general, we can write the amplitude in the form
\begin{equation}
  \mathcal{M}=\mathcal{M}_{D^*}+\mathcal{M}_D=\epsilon^{\rho\lambda\gamma\tau}(P_\tau G_1 + q_\tau G_2) \epsilon_\rho(P)\epsilon_\lambda(q_1)\epsilon_\gamma(q_2),
\end{equation}
where $G_1$ and $G_2$ are Lorentz invariant form factors.

In the calculation we stay in the rest frame of the initial bound state and hence $P = (M,0)$. We use the following input parameters: $M_{D^{*0}}$=2006.85 MeV, $M_{D^0}$=1864.80 MeV and $M_{J_\psi}$=3096.90 MeV \cite{Olive:2016xmw}. We use the numerical solution for the BS wave function to calculate the decay width of the decay $X(3872) \rightarrow$ $\gamma J/\psi$. We obtain the following decay width $\Gamma$ when $\Lambda =$ 0.69 GeV, 0.83 GeV, and 0.96 GeV, respectively:
\begin{equation}
\begin{split}
 \mathrm{when}\ \Lambda = 0.69 &\ \mathrm{GeV}, \quad \Gamma(X(3872) \rightarrow \gamma J/\psi) = 12.1 \ \mathrm{KeV},\\
 \mathrm{when}\ \Lambda = 0.83\, &\ \mathrm{GeV}, \quad \Gamma(X(3872) \rightarrow \gamma J/\psi) = 15.4 \ \mathrm{KeV},\\
 \mathrm{when}\ \Lambda = 0.96\, &\ \mathrm{GeV}, \quad \Gamma(X(3872) \rightarrow \gamma J/\psi) = 22.3 \ \mathrm{KeV}.
 \end{split}
\end{equation}

The decay width of $X(3872) \rightarrow$ $\gamma J/\psi $ has been studied by several groups with $X(3872)$ in different structures. For comparison, these results are dispalyed in Table 1 together with ours. In the table, $X(3872)$ are considered as a $c\bar{c}$ state \cite{Barnes:2003vb,Swanson:2004pp,Wang:2010ej}, a molecule state \cite{Swanson:2004pp,Dong:2008gb,Aceti:2012cb,Badalian:2012jz}, a mixture of a charmonium with a $D^0 \bar{D}^{*0}/\bar{D}^0 D^{*0}$ component \cite{Dong:2009uf,Cardoso:2014xda}, a tetraquark state ($c\bar{c}q\bar{q}$) \cite{Dubnicka:2011mm}, and a mixture of $c\bar{c}$ and $c\bar{c}q\bar{q}$ state \cite{Nielsen:2010ij}, respectively.

\begin{table}
  Table 1 The decay widths (in KeV) of $X(3872)\rightarrow \gamma J/\psi$ in various theoretical approaches.
  \begin{tabular}{lllllllllllll}
     \hline
                    Ref.         & \cite{Barnes:2003vb} & \cite{Swanson:2004pp} & \cite{Wang:2010ej} & \cite{Swanson:2004pp} & \cite{Dong:2008gb} & \cite{Aceti:2012cb} & \cite{Badalian:2012jz} & \cite{Dong:2009uf}& \cite{Cardoso:2014xda} & \cite{Dubnicka:2011mm}& \cite{Nielsen:2010ij} & Our result \\
                    \hline
     $\Gamma_{J/\psi \gamma}$ &         11.0         &         71-139        &         33         &        8              &      124.8-251.4   &         117      &76.6    &      1.94-16.8      &     24.7 &  10 &  1.8 $\times 10^3 $  & 12.1-22.3  \\
     \hline
   \end{tabular}
\end{table}

\section{summary and conclusion}
\label{sum-con}

In this paper, in order to investigate the structure of the observed state $X(3872)$ with the quantum numbers $J^P = 1^{++}$, we use the BS equation which has been successfully applied in many theoretical studies concerning heavy mesons and baryons and automatically includes relativistic corrections. We work in the picture that $X(3872)$ is an $S$-wave $D\bar{D}^*$ molecular bound state because it is very close to the $D\bar{D}^*$ threshold. We establish the BS equation for the system composed of a vector meson and a pseudoscalar meson. Then we derive the BS equation for the $D\bar{D}^*$ system using the kernel which is induced by $\sigma$, $\pi$, $\eta$, $\rho$ and $\omega$ exchange diagrams. In our model, we have used the ladder approximation which can considerably simplify the formalism. In addition, based on the fact that the $D\bar{D}^*$ system is  very weekly bound, we have used the instantaneous approximation in the BS equation, in which the energy exchange between the constituent particles is neglected. Since the constituent particles and the exchanged particles in the $D\bar{D}^*$ system are not pointlike, we introduce form factors including a cutoff $\Lambda$ which reflects the effects of structure of these particles. Since $\Lambda$ is controlled by nonperturbative QCD and cannot be determined at present, we let it vary in a reasonable range to find its values with which $X(3872)$ can be a $D\bar{D}^*$ molecular bound state.

From the calculating results we find that $X(3872)$ can be a $D\bar{D}^*$ molecular bound state. Unfortunately, we cannot determine the binding energy uniquely. The binding energy depends on the value of the cutoff $\Lambda$. We find cutoff values for which the solutions (with the binding energy $E_b \in $ (-0.01, -0.9) MeV which effect on the BS wave function is negligible) to the $X(3872)$ of the BS equation can be found: $\Lambda$ = 0.69-0.96 GeV.

We apply the numerical solution for the BS wave function to calculate the decay width $X(3872) \rightarrow J/\psi \gamma$ which is induced by $D$ and $D^*$ exchange diagrams. We predict that the decay width of $X(3872) \rightarrow J/\psi \gamma$ is in the range 12.1-22.3 KeV when $\Lambda$ varies in the range 0.69-0.96 GeV.

\acknowledgments
This work was supported by National Natural Science Foundation of China (Projects No. 11275025, No. 11575023 and No. 11775024) and the Fundamental Research Funds for the Central Universities
of China (Project No. 31020170QD052).

\begin{appendix}

\section{The expressions of $V_1(|p_t|,|q_t|)$ and $V_2(|p_t|,|q_t|)$}
\label{appendix}
\begin{tiny}
\begin{equation}
\begin{split}
&V_1(|p_t|,|q_t|)=-\frac{|q_t|}{48 |p_t| \omega_1 (M+\omega_1-\omega_2) (M+\omega_1+\omega_2)}\Bigg\{
 -\frac{48g_{DD\mathbb{V}}g_{D^*D^*\mathbb{V}}|p_t|^3|q_t|\left(\Lambda^2-m_\omega^2\right)\left[\Lambda^2+2\left(|p_t|^2+|q_t|^2-2M\omega_1-2\omega_1^2\right)\right]}
 {m_1^2\left[\Lambda^2+\left(|p_t|-|q_t|\right)^2\right]\left[\Lambda^2+\left(|p_t|+|q_t|\right)^2\right]}\\
&-\frac{48g_{DD\mathbb{V}}g_{D^*D^*\mathbb{V}}|p_t|^3|q_t|\left(\Lambda^2-m_\rho^2\right)\left[\Lambda^2+2\left(|p_t|^2+|q_t|^2-2M\omega_1-2\omega_1^2\right)\right]}
 {m_1^2\left[\Lambda^2+\left(|p_t|-|q_t|\right)^2\right]\left[\Lambda^2+\left(|p_t|+|q_t|\right)^2\right]}\\
&+\frac{12g_{DD\mathbb{V}}g_{D^*D^*\mathbb{V}}\left(|p_t|^2-|q_t|^2\right)^2|p_t|^2}{m_1^2m_\omega^2}\ln\frac{\Lambda^2+(|p_t|-|q_t|)^2}{\Lambda^2+(|p_t|+|q_t|)^2}
 +\frac{12g_{DD\mathbb{V}}g_{D^*D^*\mathbb{V}}\left(|p_t|^2-|q_t|^2\right)^2|p_t|^2}{m_1^2m_\rho^2}\ln\frac{\Lambda^2+(|p_t|-|q_t|)^2}{\Lambda^2+(|p_t|+|q_t|)^2}\\
&+\frac{48g_\sigma^2m_2|p_t|^2}{m_1}\ln\frac{\Lambda^2+(|p_t|-|q_t|)^2}{\Lambda^2+(|p_t|+|q_t|)^2}
 +\frac{12g_{DD\mathbb{V}}g_{D^*D^*\mathbb{V}}\left[m_\omega^2+2(|p_t|^2+|q_t|^2-2M\omega_1-2\omega_1^2)\right]|p_t|^2}{m_1^2}\ln\frac{\Lambda^2+(|p_t|-|q_t|)^2}{\Lambda^2+(|p_t|+|q_t|)^2}\\
&+\frac{12g_{DD\mathbb{V}}g_{D^*D^*\mathbb{V}}\left[m_\rho^2+2(|p_t|^2+|q_t|^2-2M\omega_1-2\omega_1^2)\right]|p_t|^2}{m_1^2}\ln\frac{\Lambda^2+(|p_t|-|q_t|)^2}{\Lambda^2+(|p_t|+|q_t|)^2}
 +\frac{12g_{DD\mathbb{V}}g_{D^*D^*\mathbb{V}}\left(|p_t|^2-|q_t|^2\right)^2|p_t|^2}{m_1^2m_\omega^2}\ln\frac{m_\omega^2+(|p_t|-|q_t|)^2}{m_\omega^2+(|p_t|+|q_t|)^2}\\
&+\frac{12g_{DD\mathbb{V}}g_{D^*D^*\mathbb{V}}\left[m_\omega^2+2(|p_t|^2+|q_t|^2-2M\omega_1-2\omega_1^2)\right]|p_t|^2}{m_1^2}\ln\frac{m_\omega^2+(|p_t|-|q_t|)^2}{m_\omega^2+(|p_t|+|q_t|)^2}
 +\frac{12g_{DD\mathbb{V}}g_{D^*D^*\mathbb{V}}\left(|p_t|^2-|q_t|^2\right)^2|p_t|^2}{m_1^2m_\rho^2}\ln\frac{m_\rho^2+(|p_t|-|q_t|)^2}{m_\rho^2+(|p_t|+|q_t|)^2}\\
&+\frac{12g_{DD\mathbb{V}}g_{D^*D^*\mathbb{V}}\left[m_\rho^2+2(|p_t|^2+|q_t|^2-2M\omega_1-2\omega_1^2)\right]|p_t|^2}{m_1^2}\ln\frac{m_\rho^2+(|p_t|-|q_t|)^2}{m_\rho^2+(|p_t|+|q_t|)^2}
 +\frac{48g_\sigma^2m_2|p_t|^2}{m_1}\ln\frac{m_\sigma^2+(|p_t|-|q_t|)^2}{m_\sigma^2+(|p_t|+|q_t|)^2}\\
&-\frac{48g_{DD\mathbb{V}}g_{D^*D^*\mathbb{V}}|p_t|^3|q_t|\left(\Lambda^2-m_\omega^2\right)\left(|p_t|^2-|q_t|^2\right)^2}{m_1^2m_\omega^2\left[\Lambda^2+(|p_t|-|q_t|)^2\right]\left[\Lambda^2+(|p_t|+|q_t|)^2\right]}
 -\frac{48g_{DD\mathbb{V}}g_{D^*D^*\mathbb{V}}|p_t|^3|q_t|\left(\Lambda^2-m_\rho^2\right)\left(|p_t|^2-|q_t|^2\right)^2}{m_1^2m_\rho^2\left[\Lambda^2+(|p_t|-|q_t|)^2\right]\left[\Lambda^2+(|p_t|+|q_t|)^2\right]}
 +\frac{192g_\sigma^2m_2|p_t|^3|q_t|\left(\Lambda^2-m_\sigma^2\right)}{m_1\left[\Lambda^2+(|p_t|-|q_t|)^2\right]\left[\Lambda^2+(|p_t|+|q_t|)^2\right]}\\
&+\frac{48g_{DD\mathbb{V}}f_{D^*D^*\mathbb{V}}|p_t||q_t|\left(\Lambda^2-m_\omega^2\right)\left[\Lambda^4+2\Lambda^2\left(|p_t|^2+|q_t|^2\right)-3|p_t|^4+|q_t|^4+2|p_t|^2|q_t|^2\right]}{m_1^2\left[\Lambda^2+\left(|p_t|-|q_t|\right)^2\right]\left[\Lambda^2+\left(|p_t|+|q_t|\right)^2\right]}\\
&+\frac{48g_{DD\mathbb{V}}f_{D^*D^*\mathbb{V}}|p_t||q_t|\left(\Lambda^2-m_\rho^2\right)\left[\Lambda^4+2\Lambda^2\left(|p_t|^2+|q_t|^2\right)-3|p_t|^4+|q_t|^4+2|p_t|^2|q_t|^2\right]}{m_1^2\left[\Lambda^2+\left(|p_t|-|q_t|\right)^2\right]\left[\Lambda^2+\left(|p_t|+|q_t|\right)^2\right]}\\
&+\frac{48f_{DD^*\mathbb{V}}^2M^2|p_t||q_t|\left(\Lambda^2-m_\omega^2\right)\left[\Lambda^4+2\Lambda^2\left(|p_t|^2+|q_t|^2\right)+\left(|p_t|^2-|p_t|^2\right)^2\right]}{m_1^2\left[\Lambda^2+\left(|p_t|-|q_t|\right)^2\right]\left[\Lambda^2+\left(|p_t|+|q_t|\right)^2\right]}
 +\frac{48f_{DD^*\mathbb{V}}^2M^2|p_t||q_t|\left(\Lambda^2-m_\rho^2\right)\left[\Lambda^4+2\Lambda^2\left(|p_t|^2+|q_t|^2\right)+\left(|p_t|^2-|p_t|^2\right)^2\right]}{m_1^2\left[\Lambda^2+\left(|p_t|-|q_t|\right)^2\right]\left[\Lambda^2+\left(|p_t|+|q_t|\right)^2\right]}\\
&-\frac{48g_{DD\mathbb{V}}f_{D^*D^*\mathbb{V}}|p_t||q_t|\left(\Lambda^2-m_\omega^2\right)\left(\Lambda^2-|p_t|^2+|q_t|^2\right)\left(\Lambda^2+3|p_t|^2+|q_t|^2-4\omega_1^2-4M\omega_1\right)}{m_1^2\left[\Lambda^2+\left(|p_t|-|q_t|\right)^2\right]\left[\Lambda^2+\left(|p_t|+|q_t|\right)^2\right]}\\
&-\frac{48g_{DD\mathbb{V}}f_{D^*D^*\mathbb{V}}|p_t||q_t|\left(\Lambda^2-m_\rho^2\right)\left(\Lambda^2-|p_t|^2+|q_t|^2\right)\left(\Lambda^2+3|p_t|^2+|q_t|^2-4\omega_1^2-4M\omega_1\right)}{m_1^2\left[\Lambda^2+\left(|p_t|-|q_t|\right)^2\right]\left[\Lambda^2+\left(|p_t|+|q_t|\right)^2\right]}\\
&+\frac{48g_{DD\mathbb{V}}g_{D^*D^*\mathbb{V}}|p_t||q_t|\left(\Lambda^2-m_\omega^2\right)\left[\Lambda^2+2\left(|p_t|^2+|q_t|^2-2M\omega_1-2\omega_1^2\right)\right]}{\left[\Lambda^2+\left(|p_t|-|q_t|\right)^2\right]\left[\Lambda^2+\left(|p_t|+|q_t|\right)^2\right]}
 +\frac{48g_{DD\mathbb{V}}g_{D^*D^*\mathbb{V}}|p_t||q_t|\left(\Lambda^2-m_\rho^2\right)\left[\Lambda^2+2\left(|p_t|^2+|q_t|^2-2M\omega_1-2\omega_1^2\right)\right]}{\left[\Lambda^2+\left(|p_t|-|q_t|\right)^2\right]\left[\Lambda^2+\left(|p_t|+|q_t|\right)^2\right]}\\
&+48g_\sigma^2m_1m_2\ln\frac{\Lambda^2+(|p_t|-|q_t|)^2}{\Lambda^2+(|p_t|+|q_t|)^2}
 +\frac{g_{DD^*\mathbb{P}}^2\left(\Lambda^2-2m_\eta^2+|p_t|^2-|q_t|^2\right)\left(\Lambda^2-|p_t|^2+|q_t|^2\right)}{m_1^2}\ln\frac{\Lambda^2+(|p_t|-|q_t|)^2}{\Lambda^2+(|p_t|+|q_t|)^2}\\
&+\frac{3g_{DD^*\mathbb{P}}^2\left(\Lambda^2-2m_\pi^2+|p_t|^2-|q_t|^2\right)\left(\Lambda^2-|p_t|^2+|q_t|^2\right)}{m_1^2}\ln\frac{\Lambda^2+(|p_t|-|q_t|)^2}{\Lambda^2+(|p_t|+|q_t|)^2}\\
&+\frac{12g_{DD\mathbb{V}}f_{D^*D^*\mathbb{V}}\left[\Lambda^4-2\Lambda^2m_\omega^2+3|p_t|^4-|q_t|^4-2|p_t|^2|q_t|^2-2m_\omega^2\left(|p_t|^2+|q_t|^2\right)\right]}{m_1^2}\ln\frac{\Lambda^2+(|p_t|-|q_t|)^2}{\Lambda^2+(|p_t|+|q_t|)^2}\\
&+\frac{12f_{DD^*\mathbb{V}}^2M^2\left[-\Lambda^4+2\Lambda^2m_\omega^2+\left(|p_t|^2-|q_t|^2\right)^2+2m_\omega^2\left(|p_t|^2+|q_t|^2\right)\right]}{m_1^2}\ln\frac{\Lambda^2+(|p_t|-|q_t|)^2}{\Lambda^2+(|p_t|+|q_t|)^2}\\
&+\frac{12g_{DD\mathbb{V}}f_{D^*D^*\mathbb{V}}\left[\Lambda^4-2\Lambda^2m_\rho^2+3|p_t|^4-|q_t|^4-2|p_t|^2|q_t|^2-2m_\rho^2\left(|p_t|^2+|q_t|^2\right)\right]}{m_1^2}\ln\frac{\Lambda^2+(|p_t|-|q_t|)^2}{\Lambda^2+(|p_t|+|q_t|)^2}\\
&+\frac{12f_{DD^*\mathbb{V}}^2M^2\left[-\Lambda^4+2\Lambda^2m_\rho^2+\left(|p_t|^2-|q_t|^2\right)^2+2m_\rho^2\left(|p_t|^2+|q_t|^2\right)\right]}{m_1^2}\ln\frac{\Lambda^2+(|p_t|-|q_t|)^2}{\Lambda^2+(|p_t|+|q_t|)^2}\\
&+12g_{DD\mathbb{V}}g_{D^*D^*\mathbb{V}}\left[m_\omega^2+2\left(|p_t|^2+|q_t|^2-2M\omega_1-2\omega_1^2\right)\right]\ln\frac{\Lambda^2+(|p_t|-|q_t|)^2}{\Lambda^2+(|p_t|+|q_t|)^2}\\
&+12g_{DD\mathbb{V}}g_{D^*D^*\mathbb{V}}\left[m_\rho^2+2\left(|p_t|^2+|q_t|^2-2M\omega_1-2\omega_1^2\right)\right]\ln\frac{\Lambda^2+(|p_t|-|q_t|)^2}{\Lambda^2+(|p_t|+|q_t|)^2}\\
&+\frac{12g_{DD\mathbb{V}}f_{D^*D^*\mathbb{V}}\left[\Lambda^4-2\Lambda^2m_\omega^2+\left(|p_t|^2-|q_t|^2\right)\left(3|p_t|^2+|q_t|^2-4M\omega_1-4\omega_1^2\right)-2m_\omega^2\left(|p_t|^2+|q_t|^2-2M\omega_1-2\omega_1^2\right)\right]}{m_1^2}\ln\frac{\Lambda^2+(|p_t|-|q_t|)^2}{\Lambda^2+(|p_t|+|q_t|)^2}\\
&+\frac{12g_{DD\mathbb{V}}f_{D^*D^*\mathbb{V}}\left[\Lambda^4-2\Lambda^2m_\rho^2+\left(|p_t|^2-|q_t|^2\right)\left(3|p_t|^2+|q_t|^2-4M\omega_1-4\omega_1^2\right)-2m_\rho^2\left(|p_t|^2+|q_t|^2-2M\omega_1-2\omega_1^2\right)\right]}{m_1^2}\ln\frac{\Lambda^2+(|p_t|-|q_t|)^2}{\Lambda^2+(|p_t|+|q_t|)^2}\\
&+\frac{g_{DD^*\mathbb{P}}^2\left(m_\eta^2-|p_t|^2+|q_t|^2\right)^2}{m_1^2}\ln\frac{m_\eta^2+(|p_t|-|q_t|)^2}{m_\eta^2+(|p_t|+|q_t|)^2}
 +\frac{12g_{DD\mathbb{V}}f_{D^*D^*\mathbb{V}}\left[m_\omega^4+2m_\omega^2\left(|p_t|^2-|q_t|^2\right)-3|p_t|^4+|q_t|^4+2|p_t|^2|q_t|^2\right]}{m_1^2}\ln\frac{m_\omega^2+\left(|p_t|-|q_t|\right)^2}{m_\omega^2+\left(|p_t|+|q_t|\right)^2}\\
&+\frac{12f_{DD^*\mathbb{V}}^2M^2\left[m_\omega^4+2m_\omega^2\left(|p_t|^2+|q_t|^2\right)+\left(|p_t|^2-|q_t|^2\right)^2\right]}{m_1^2}\ln\frac{m_\omega^2+\left(|p_t|-|q_t|\right)^2}{m_\omega^2+\left(|p_t|+|q_t|\right)^2}\\
&+\frac{12g_{DD\mathbb{V}}f_{D^*D^*\mathbb{V}}\left(m_\omega^2-|p_t|^2+|q_t|^2\right)\left(m_\omega^2+3|p_t|^2+|q_t|^2-4M\omega_1-4\omega_1^2\right)}{m_1^2}\ln\frac{m_\omega^2+\left(|p_t|-|q_t|\right)^2}{m_\omega^2+\left(|p_t|+|q_t|\right)^2}\\
\end{split}
\end{equation}
\end{tiny}

\begin{tiny}
\begin{equation}\nonumber
\begin{split}
&+12g_{DD\mathbb{V}}g_{D^*D^*\mathbb{V}}\left[m_\omega^2+2\left(|p_t|^2+|q_t|^2-2M\omega_1-2\omega_1^2\right)\right]\ln\frac{m_\omega^2+\left(|p_t|-|q_t|\right)^2}{m_\omega^2+\left(|p_t|+|q_t|\right)^2}
 +\frac{3g_{DD^*\mathbb{P}}^2\left(m_\pi^2-|p_t|^2+|q_t|^2\right)^2}{m_1^2}\ln\frac{m_\pi^2+\left(|p_t|-|q_t|\right)^2}{m_\pi^2+\left(|p_t|+|q_t|\right)^2}\\
&+\frac{12g_{DD\mathbb{V}}f_{D^*D^*\mathbb{V}}\left[m_\rho^4+2m_\rho^2\left(|p_t|^2-|q_t|^2\right)-3|p_t|^4+|q_t|^4+2|p_t|^2|q_t|^2\right]}{m_1^2}\ln\frac{m_\rho^2+\left(|p_t|-|q_t|\right)^2}{m_\rho^2+\left(|p_t|+|q_t|\right)^2}\\
&+\frac{12f_{DD^*\mathbb{V}}^2M^2\left[m_\rho^4+2m_\rho^2\left(|p_t|^2+|q_t|^2\right)+\left(|p_t|^2-|q_t|^2\right)^2\right]}{m_1^2}\ln\frac{m_\rho^2+\left(|p_t|-|q_t|\right)^2}{m_\rho^2+\left(|p_t|+|q_t|\right)^2}\\
&+\frac{12g_{DD\mathbb{V}}f_{D^*D^*\mathbb{V}}\left(m_\rho^2-|p_t|^2+|q_t|^2\right)\left(m_\rho^2+3|p_t|^2+|q_t|^2-4M\omega_1-4\omega_1^2\right)}{m_1^2}\ln\frac{m_\rho^2+\left(|p_t|-|q_t|\right)^2}{m_\rho^2+\left(|p_t|+|q_t|\right)^2}\\
&+12g_{DD\mathbb{V}}g_{D^*D^*\mathbb{V}}\left[m_\rho^2+2\left(|p_t|^2+|q_t|^2-2M\omega_1-2\omega_1^2\right)\right]\ln\frac{m_\rho^2+\left(|p_t|-|q_t|\right)^2}{m_\rho^2+\left(|p_t|+|q_t|\right)^2}\\
&+48g_\sigma^2m_1m_2\ln\frac{m_\sigma^2+(|p_t|-|q_t|)^2}{m_\sigma^2+(|p_t|+|q_t|)^2}
 +\frac{4g_{DD^*\mathbb{P}}^2|p_t||q_t|\left(\Lambda^2-m_\eta^2\right)\left(\Lambda^2-|p_t|^2+|q_t|^2\right)^2}{m_1^2\left[\Lambda^2+\left(|p_t|-|q_t|\right)^2\right]\left[\Lambda^2+\left(|p_t|+|q_t|\right)^2\right]}\\
&+\frac{12g_{DD^*\mathbb{P}}^2|p_t||q_t|\left(\Lambda^2-m_\pi^2\right)\left(\Lambda^2-|p_t|^2+|q_t|^2\right)^2}{m_1^2\left[\Lambda^2+\left(|p_t|-|q_t|\right)^2\right]\left[\Lambda^2+\left(|p_t|+|q_t|\right)^2\right]}
 +\frac{192g_\sigma^2m_1m_2|p_t||q_t|\left(\Lambda^2-m_\sigma^2\right)}{\left[\Lambda^2+\left(|p_t|-|q_t|\right)^2\right]\left[\Lambda^2+\left(|p_t|+|q_t|\right)^2\right]}\\
&-\frac{48g_{DD\mathbb{V}}g_{D^*D^*\mathbb{V}}|p_t||q_t|\left(\Lambda^2-m_\omega^2\right)\left(|p_t|^2-|q_t|^2\right)^2}{\left[\Lambda^2+\left(|p_t|-|q_t|\right)^2\right]\left[\Lambda^2+\left(|p_t|+|q_t|\right)^2\right]}
 -\frac{48g_{DD\mathbb{V}}g_{D^*D^*\mathbb{V}}\left(\Lambda^2-m_\rho^2\right)\left(|p_t|^2-|q_t|^2\right)^2}{\left[\Lambda^2+\left(|p_t|-|q_t|\right)^2\right]\left[\Lambda^2+\left(|p_t|+|q_t|\right)^2\right]}\\
&+\frac{48f_{DD^*\mathbb{V}}^2|p_t||q_t|\left(\Lambda^2-m_\omega^2\right)\left[\Lambda^4\left(M^2+8|p_t|^2\right)+2M^2\Lambda^2\left(|q_t|^2-3|p_t|^2\right)+M^2\left(|p_t|^2-|q_t|^2\right)^2\right]}{\left[\Lambda^2+\left(|p_t|-|q_t|\right)^2\right]\left[\Lambda^2+\left(|p_t|+|q_t|\right)^2\right]}\\
&+\frac{48f_{DD^*\mathbb{V}}^2|p_t||q_t|\left(\Lambda^2-m_\rho^2\right)\left[\Lambda^4\left(M^2+8|p_t|^2\right)+2M^2\Lambda^2\left(|q_t|^2-3|p_t|^2\right)+M^2\left(|p_t|^2-|q_t|^2\right)^2\right]}{\left[\Lambda^2+\left(|p_t|-|q_t|\right)^2\right]\left[\Lambda^2+\left(|p_t|+|q_t|\right)^2\right]}\\
&+24g_{DD\mathbb{V}}g_{D^*D^*\mathbb{V}}\left(|p_t|^2-|q_t|^2\right)^2\ln\frac{\Lambda^2+(|p_t|-|q_t|)^2}{\Lambda^2+(|p_t|+|q_t|)^2}
 +\frac{g_{DD^*\mathbb{P}}^2\left(\Lambda^2-2m_\eta^2+|p_t|^2-|q_t|^2\right)\left(\Lambda^2-|p_t|^2+|q_t|^2\right)}{|p_t|^2}\ln\frac{\Lambda^2+(|p_t|-|q_t|)^2}{\Lambda^2+(|p_t|+|q_t|)^2}\\
&+\frac{3g_{DD^*\mathbb{P}}^2\left(\Lambda^2-2m_\pi^2+|p_t|^2-|q_t|^2\right)\left(\Lambda^2-|p_t|^2+|q_t|^2\right)}{|p_t|^2}\ln\frac{\Lambda^2+(|p_t|-|q_t|)^2}{\Lambda^2+(|p_t|+|q_t|)^2}\\
&-\frac{12f_{DD^*\mathbb{V}}^2\left[-\Lambda^4\left(M^2+8|p_t|^2\right)+2\Lambda^2m_\omega^2\left(M^2+8|p_t|^2\right)+M^2m_\omega^2\left(2|q_t|^2-6|p_t|^2\right)+M^2\left(|p_t|^2-|q_t|^2\right)^2\right]}{|p_t|^2}\ln\frac{\Lambda^2+(|p_t|-|q_t|)^2}{\Lambda^2+(|p_t|+|q_t|)^2}\\
&-\frac{12f_{DD^*\mathbb{V}}^2\left[-\Lambda^4\left(M^2+8|p_t|^2\right)+2\Lambda^2m_\rho^2\left(M^2+8|p_t|^2\right)+M^2m_\rho^2\left(2|q_t|^2-6|p_t|^2\right)+M^2\left(|p_t|^2-|q_t|^2\right)^2\right]}{|p_t|^2}\ln\frac{\Lambda^2+(|p_t|-|q_t|)^2}{\Lambda^2+(|p_t|+|q_t|)^2}\\
&+\frac{g_{DD^*\mathbb{P}}^2\left(m_\eta^2-|p_t|^2+|q_t|^2\right)^2}{|p_t|^2}\ln\frac{m_\eta^2+\left(|p_t|-|q_t|\right)^2}{m_\eta^2+\left(|p_t|+|q_t|\right)^2}
 -12g_{DD\mathbb{V}}g_{D^*D^*\mathbb{V}}\left(|p_t|^2-|q_t|^2\right)^2\ln\frac{m_\omega^2+\left(|p_t|-|q_t|\right)^2}{m_\omega^2+\left(|p_t|+|q_t|\right)^2}\\
&+\frac{12f_{DD^*\mathbb{V}}^2\left[8|p_t|^2m_\omega^4+M^2m_\omega^4+M^2m_\omega^2\left(2|q_t|^2-6|p_t|^2\right)+M^2\left(|p_t|^2-|q_t|^2\right)^2\right]}{|p_t|^2}\ln\frac{m_\omega^2+\left(|p_t|-|q_t|\right)^2}{m_\omega^2+\left(|p_t|+|q_t|\right)^2}
 +\frac{3g_{DD^*\mathbb{P}}^2\left(m_\pi^2-|p_t|^2+|q_t|^2\right)^2}{|p_t|^2}\ln\frac{m_\pi^2+\left(|p_t|-|q_t|\right)^2}{m_\pi^2+\left(|p_t|+|q_t|\right)^2}\\
&-12g_{DD\mathbb{V}}g_{D^*D^*\mathbb{V}}\left(|p_t|^2-|q_t|^2\right)^2\ln\frac{m_\rho^2+\left(|p_t|-|q_t|\right)^2}{m_\rho^2+\left(|p_t|+|q_t|\right)^2}
 +\frac{12f_{DD^*\mathbb{V}}^2\left[8|p_t|^2m_\rho^4+M^2m_\rho^4+M^2m_\rho^2\left(2|q_t|^2-6|p_t|^2\right)+M^2\left(|p_t|^2-|q_t|^2\right)^2\right]}{|p_t|^2}\ln\frac{m_\rho^2+\left(|p_t|-|q_t|\right)^2}{m_\rho^2+\left(|p_t|+|q_t|\right)^2}\\
&+\frac{4g_{DD^*\mathbb{P}}^2|p_t||q_t|\left(\Lambda^2-m_\eta^2\right)\left(\Lambda^2-|p_t|^2+|q_t|^2\right)^2}{|p_t|^2\left[\Lambda^2+\left(|p_t|-|q_t|\right)^2\right]\left[\Lambda^2+\left(|p_t|+|q_t|\right)^2\right]}
 +\frac{12g_{DD^*\mathbb{P}}^2|p_t||q_t|\left(\Lambda^2-m_\pi^2\right)\left(\Lambda^2-|p_t|^2+|q_t|^2\right)^2}{|p_t|^2\left[\Lambda^2+\left(|p_t|-|q_t|\right)^2\right]\left[\Lambda^2+\left(|p_t|+|q_t|\right)^2\right]}\Bigg\},
\end{split}
\end{equation}
\end{tiny}

\begin{tiny}
\begin{equation}
\begin{split}
&V_2(|p_t|,|q_t|)=-\frac{|q_t|}{48 |p_t| \omega_2 (M-\omega_1+\omega_2) (M+\omega_1+\omega_2)}\Bigg\{
 -\frac{48g_{DD\mathbb{V}}g_{D^*D^*\mathbb{V}}|p_t|^3|q_t|\left(\Lambda^2-m_\omega^2\right)\left[\Lambda^2+2\left(|p_t|^2+|q_t|^2-2M\omega_2-2\omega_2^2\right)\right]}
 {m_1^2\left[\Lambda^2+\left(|p_t|-|q_t|\right)^2\right]\left[\Lambda^2+\left(|p_t|+|q_t|\right)^2\right]}\\
&-\frac{48g_{DD\mathbb{V}}g_{D^*D^*\mathbb{V}}|p_t|^3|q_t|\left(\Lambda^2-m_\rho^2\right)\left[\Lambda^2+2\left(|p_t|^2+|q_t|^2-2M\omega_2-2\omega_2^2\right)\right]}
 {m_1^2\left[\Lambda^2+\left(|p_t|-|q_t|\right)^2\right]\left[\Lambda^2+\left(|p_t|+|q_t|\right)^2\right]}\\
&+\frac{12g_{DD\mathbb{V}}g_{D^*D^*\mathbb{V}}\left(|p_t|^2-|q_t|^2\right)^2|p_t|^2}{m_1^2m_\omega^2}\ln\frac{\Lambda^2+(|p_t|-|q_t|)^2}{\Lambda^2+(|p_t|+|q_t|)^2}
 +\frac{12g_{DD\mathbb{V}}g_{D^*D^*\mathbb{V}}\left(|p_t|^2-|q_t|^2\right)^2|p_t|^2}{m_1^2m_\rho^2}\ln\frac{\Lambda^2+(|p_t|-|q_t|)^2}{\Lambda^2+(|p_t|+|q_t|)^2}\\
&+\frac{48g_\sigma^2m_2|p_t|^2}{m_1}\ln\frac{\Lambda^2+(|p_t|-|q_t|)^2}{\Lambda^2+(|p_t|+|q_t|)^2}
 +\frac{12g_{DD\mathbb{V}}g_{D^*D^*\mathbb{V}}\left[m_\omega^2+2(|p_t|^2+|q_t|^2-2M\omega_2-2\omega_2^2)\right]|p_t|^2}{m_1^2}\ln\frac{\Lambda^2+(|p_t|-|q_t|)^2}{\Lambda^2+(|p_t|+|q_t|)^2}\\
&+\frac{12g_{DD\mathbb{V}}g_{D^*D^*\mathbb{V}}\left[m_\rho^2+2(|p_t|^2+|q_t|^2-2M\omega_2-2\omega_2^2)\right]|p_t|^2}{m_1^2}\ln\frac{\Lambda^2+(|p_t|-|q_t|)^2}{\Lambda^2+(|p_t|+|q_t|)^2}
 +\frac{12g_{DD\mathbb{V}}g_{D^*D^*\mathbb{V}}\left(|p_t|^2-|q_t|^2\right)^2|p_t|^2}{m_1^2m_\omega^2}\ln\frac{m_\omega^2+(|p_t|-|q_t|)^2}{m_\omega^2+(|p_t|+|q_t|)^2}\\
&+\frac{12g_{DD\mathbb{V}}g_{D^*D^*\mathbb{V}}\left[m_\omega^2+2(|p_t|^2+|q_t|^2-2M\omega_2-2\omega_2^2)\right]|p_t|^2}{m_1^2}\ln\frac{m_\omega^2+(|p_t|-|q_t|)^2}{m_\omega^2+(|p_t|+|q_t|)^2}
 +\frac{12g_{DD\mathbb{V}}g_{D^*D^*\mathbb{V}}\left(|p_t|^2-|q_t|^2\right)^2|p_t|^2}{m_1^2m_\rho^2}\ln\frac{m_\rho^2+(|p_t|-|q_t|)^2}{m_\rho^2+(|p_t|+|q_t|)^2}\\
&+\frac{12g_{DD\mathbb{V}}g_{D^*D^*\mathbb{V}}\left[m_\rho^2+2(|p_t|^2+|q_t|^2-2M\omega_2-2\omega_2^2)\right]|p_t|^2}{m_1^2}\ln\frac{m_\rho^2+(|p_t|-|q_t|)^2}{m_\rho^2+(|p_t|+|q_t|)^2}
 +\frac{48g_\sigma^2m_2|p_t|^2}{m_1}\ln\frac{m_\sigma^2+(|p_t|-|q_t|)^2}{m_\sigma^2+(|p_t|+|q_t|)^2}\\
&-\frac{48g_{DD\mathbb{V}}g_{D^*D^*\mathbb{V}}|p_t|^3|q_t|\left(\Lambda^2-m_\omega^2\right)\left(|p_t|^2-|q_t|^2\right)^2}{m_1^2m_\omega^2\left[\Lambda^2+(|p_t|-|q_t|)^2\right]\left[\Lambda^2+(|p_t|+|q_t|)^2\right]}
 -\frac{48g_{DD\mathbb{V}}g_{D^*D^*\mathbb{V}}|p_t|^3|q_t|\left(\Lambda^2-m_\rho^2\right)\left(|p_t|^2-|q_t|^2\right)^2}{m_1^2m_\rho^2\left[\Lambda^2+(|p_t|-|q_t|)^2\right]\left[\Lambda^2+(|p_t|+|q_t|)^2\right]}
 +\frac{192g_\sigma^2m_2|p_t|^3|q_t|\left(\Lambda^2-m_\sigma^2\right)}{m_1\left[\Lambda^2+(|p_t|-|q_t|)^2\right]\left[\Lambda^2+(|p_t|+|q_t|)^2\right]}\\
&+\frac{48g_{DD\mathbb{V}}f_{D^*D^*\mathbb{V}}|p_t||q_t|\left(\Lambda^2-m_\omega^2\right)\left[\Lambda^4+2\Lambda^2\left(|p_t|^2+|q_t|^2\right)-3|p_t|^4+|q_t|^4+2|p_t|^2|q_t|^2\right]}{m_1^2\left[\Lambda^2+\left(|p_t|-|q_t|\right)^2\right]\left[\Lambda^2+\left(|p_t|+|q_t|\right)^2\right]}\\
&+\frac{48g_{DD\mathbb{V}}f_{D^*D^*\mathbb{V}}|p_t||q_t|\left(\Lambda^2-m_\rho^2\right)\left[\Lambda^4+2\Lambda^2\left(|p_t|^2+|q_t|^2\right)-3|p_t|^4+|q_t|^4+2|p_t|^2|q_t|^2\right]}{m_1^2\left[\Lambda^2+\left(|p_t|-|q_t|\right)^2\right]\left[\Lambda^2+\left(|p_t|+|q_t|\right)^2\right]}\\
&+\frac{48f_{DD^*\mathbb{V}}^2M^2|p_t||q_t|\left(\Lambda^2-m_\omega^2\right)\left[\Lambda^4+2\Lambda^2\left(|p_t|^2+|q_t|^2\right)+\left(|p_t|^2-|p_t|^2\right)^2\right]}{m_1^2\left[\Lambda^2+\left(|p_t|-|q_t|\right)^2\right]\left[\Lambda^2+\left(|p_t|+|q_t|\right)^2\right]}
 +\frac{48f_{DD^*\mathbb{V}}^2M^2|p_t||q_t|\left(\Lambda^2-m_\rho^2\right)\left[\Lambda^4+2\Lambda^2\left(|p_t|^2+|q_t|^2\right)+\left(|p_t|^2-|p_t|^2\right)^2\right]}{m_1^2\left[\Lambda^2+\left(|p_t|-|q_t|\right)^2\right]\left[\Lambda^2+\left(|p_t|+|q_t|\right)^2\right]}\\
&-\frac{48g_{DD\mathbb{V}}f_{D^*D^*\mathbb{V}}|p_t||q_t|\left(\Lambda^2-m_\omega^2\right)\left(\Lambda^2-|p_t|^2+|q_t|^2\right)\left(\Lambda^2+3|p_t|^2+|q_t|^2-4\omega_2^2-4M\omega_2\right)}{m_1^2\left[\Lambda^2+\left(|p_t|-|q_t|\right)^2\right]\left[\Lambda^2+\left(|p_t|+|q_t|\right)^2\right]}\\
&-\frac{48g_{DD\mathbb{V}}f_{D^*D^*\mathbb{V}}|p_t||q_t|\left(\Lambda^2-m_\rho^2\right)\left(\Lambda^2-|p_t|^2+|q_t|^2\right)\left(\Lambda^2+3|p_t|^2+|q_t|^2-4\omega_2^2-4M\omega_2\right)}{m_1^2\left[\Lambda^2+\left(|p_t|-|q_t|\right)^2\right]\left[\Lambda^2+\left(|p_t|+|q_t|\right)^2\right]}\\
&+\frac{48g_{DD\mathbb{V}}g_{D^*D^*\mathbb{V}}|p_t||q_t|\left(\Lambda^2-m_\omega^2\right)\left[\Lambda^2+2\left(|p_t|^2+|q_t|^2-2M\omega_2-2\omega_2^2\right)\right]}{\left[\Lambda^2+\left(|p_t|-|q_t|\right)^2\right]\left[\Lambda^2+\left(|p_t|+|q_t|\right)^2\right]}
 +\frac{48g_{DD\mathbb{V}}g_{D^*D^*\mathbb{V}}|p_t||q_t|\left(\Lambda^2-m_\rho^2\right)\left[\Lambda^2+2\left(|p_t|^2+|q_t|^2-2M\omega_2-2\omega_2^2\right)\right]}{\left[\Lambda^2+\left(|p_t|-|q_t|\right)^2\right]\left[\Lambda^2+\left(|p_t|+|q_t|\right)^2\right]}\\
&+48g_\sigma^2m_1m_2\ln\frac{\Lambda^2+(|p_t|-|q_t|)^2}{\Lambda^2+(|p_t|+|q_t|)^2}
 +\frac{g_{DD^*\mathbb{P}}^2\left(\Lambda^2-2m_\eta^2+|p_t|^2-|q_t|^2\right)\left(\Lambda^2-|p_t|^2+|q_t|^2\right)}{m_1^2}\ln\frac{\Lambda^2+(|p_t|-|q_t|)^2}{\Lambda^2+(|p_t|+|q_t|)^2}\\
&+\frac{3g_{DD^*\mathbb{P}}^2\left(\Lambda^2-2m_\pi^2+|p_t|^2-|q_t|^2\right)\left(\Lambda^2-|p_t|^2+|q_t|^2\right)}{m_1^2}\ln\frac{\Lambda^2+(|p_t|-|q_t|)^2}{\Lambda^2+(|p_t|+|q_t|)^2}\\
&+\frac{12g_{DD\mathbb{V}}f_{D^*D^*\mathbb{V}}\left[\Lambda^4-2\Lambda^2m_\omega^2+3|p_t|^4-|q_t|^4-2|p_t|^2|q_t|^2-2m_\omega^2\left(|p_t|^2+|q_t|^2\right)\right]}{m_1^2}\ln\frac{\Lambda^2+(|p_t|-|q_t|)^2}{\Lambda^2+(|p_t|+|q_t|)^2}\\
&+\frac{12f_{DD^*\mathbb{V}}^2M^2\left[-\Lambda^4+2\Lambda^2m_\omega^2+\left(|p_t|^2-|q_t|^2\right)^2+2m_\omega^2\left(|p_t|^2+|q_t|^2\right)\right]}{m_1^2}\ln\frac{\Lambda^2+(|p_t|-|q_t|)^2}{\Lambda^2+(|p_t|+|q_t|)^2}\\
&+\frac{12g_{DD\mathbb{V}}f_{D^*D^*\mathbb{V}}\left[\Lambda^4-2\Lambda^2m_\rho^2+3|p_t|^4-|q_t|^4-2|p_t|^2|q_t|^2-2m_\rho^2\left(|p_t|^2+|q_t|^2\right)\right]}{m_1^2}\ln\frac{\Lambda^2+(|p_t|-|q_t|)^2}{\Lambda^2+(|p_t|+|q_t|)^2}\\
&+\frac{12f_{DD^*\mathbb{V}}^2M^2\left[-\Lambda^4+2\Lambda^2m_\rho^2+\left(|p_t|^2-|q_t|^2\right)^2+2m_\rho^2\left(|p_t|^2+|q_t|^2\right)\right]}{m_1^2}\ln\frac{\Lambda^2+(|p_t|-|q_t|)^2}{\Lambda^2+(|p_t|+|q_t|)^2}\\
&+12g_{DD\mathbb{V}}g_{D^*D^*\mathbb{V}}\left[m_\omega^2+2\left(|p_t|^2+|q_t|^2-2M\omega_2-2\omega_2^2\right)\right]\ln\frac{\Lambda^2+(|p_t|-|q_t|)^2}{\Lambda^2+(|p_t|+|q_t|)^2}\\
&+12g_{DD\mathbb{V}}g_{D^*D^*\mathbb{V}}\left[m_\rho^2+2\left(|p_t|^2+|q_t|^2-2M\omega_2-2\omega_2^2\right)\right]\ln\frac{\Lambda^2+(|p_t|-|q_t|)^2}{\Lambda^2+(|p_t|+|q_t|)^2}\\
&+\frac{12g_{DD\mathbb{V}}f_{D^*D^*\mathbb{V}}\left[\Lambda^4-2\Lambda^2m_\omega^2+\left(|p_t|^2-|q_t|^2\right)\left(3|p_t|^2+|q_t|^2-4M\omega_2-4\omega_2^2\right)-2m_\omega^2\left(|p_t|^2+|q_t|^2-2M\omega_2-2\omega_2^2\right)\right]}{m_1^2}\ln\frac{\Lambda^2+(|p_t|-|q_t|)^2}{\Lambda^2+(|p_t|+|q_t|)^2}\\
&+\frac{12g_{DD\mathbb{V}}f_{D^*D^*\mathbb{V}}\left[\Lambda^4-2\Lambda^2m_\rho^2+\left(|p_t|^2-|q_t|^2\right)\left(3|p_t|^2+|q_t|^2-4M\omega_2-4\omega_2^2\right)-2m_\rho^2\left(|p_t|^2+|q_t|^2-2M\omega_2-2\omega_2^2\right)\right]}{m_1^2}\ln\frac{\Lambda^2+(|p_t|-|q_t|)^2}{\Lambda^2+(|p_t|+|q_t|)^2}\\
&+\frac{g_{DD^*\mathbb{P}}^2\left(m_\eta^2-|p_t|^2+|q_t|^2\right)^2}{m_1^2}\ln\frac{m_\eta^2+(|p_t|-|q_t|)^2}{m_\eta^2+(|p_t|+|q_t|)^2}
 +\frac{12g_{DD\mathbb{V}}f_{D^*D^*\mathbb{V}}\left[m_\omega^4+2m_\omega^2\left(|p_t|^2-|q_t|^2\right)-3|p_t|^4+|q_t|^4+2|p_t|^2|q_t|^2\right]}{m_1^2}\ln\frac{m_\omega^2+\left(|p_t|-|q_t|\right)^2}{m_\omega^2+\left(|p_t|+|q_t|\right)^2}\\
&+\frac{12f_{DD^*\mathbb{V}}^2M^2\left[m_\omega^4+2m_\omega^2\left(|p_t|^2+|q_t|^2\right)+\left(|p_t|^2-|q_t|^2\right)^2\right]}{m_1^2}\ln\frac{m_\omega^2+\left(|p_t|-|q_t|\right)^2}{m_\omega^2+\left(|p_t|+|q_t|\right)^2}\\
&+\frac{12g_{DD\mathbb{V}}f_{D^*D^*\mathbb{V}}\left(m_\omega^2-|p_t|^2+|q_t|^2\right)\left(m_\omega^2+3|p_t|^2+|q_t|^2-4M\omega_2-4\omega_2^2\right)}{m_1^2}\ln\frac{m_\omega^2+\left(|p_t|-|q_t|\right)^2}{m_\omega^2+\left(|p_t|+|q_t|\right)^2}\\
&+12g_{DD\mathbb{V}}g_{D^*D^*\mathbb{V}}\left[m_\omega^2+2\left(|p_t|^2+|q_t|^2-2M\omega_2-2\omega_2^2\right)\right]\ln\frac{m_\omega^2+\left(|p_t|-|q_t|\right)^2}{m_\omega^2+\left(|p_t|+|q_t|\right)^2}
 +\frac{3g_{DD^*\mathbb{P}}^2\left(m_\pi^2-|p_t|^2+|q_t|^2\right)^2}{m_1^2}\ln\frac{m_\pi^2+\left(|p_t|-|q_t|\right)^2}{m_\pi^2+\left(|p_t|+|q_t|\right)^2}\\
\end{split}
\end{equation}
\end{tiny}

\begin{tiny}
\begin{equation}\nonumber
\begin{split}
&+\frac{12g_{DD\mathbb{V}}f_{D^*D^*\mathbb{V}}\left[m_\rho^4+2m_\rho^2\left(|p_t|^2-|q_t|^2\right)-3|p_t|^4+|q_t|^4+2|p_t|^2|q_t|^2\right]}{m_1^2}\ln\frac{m_\rho^2+\left(|p_t|-|q_t|\right)^2}{m_\rho^2+\left(|p_t|+|q_t|\right)^2}\\
&+\frac{12f_{DD^*\mathbb{V}}^2M^2\left[m_\rho^4+2m_\rho^2\left(|p_t|^2+|q_t|^2\right)+\left(|p_t|^2-|q_t|^2\right)^2\right]}{m_1^2}\ln\frac{m_\rho^2+\left(|p_t|-|q_t|\right)^2}{m_\rho^2+\left(|p_t|+|q_t|\right)^2}\\
&+\frac{12g_{DD\mathbb{V}}f_{D^*D^*\mathbb{V}}\left(m_\rho^2-|p_t|^2+|q_t|^2\right)\left(m_\rho^2+3|p_t|^2+|q_t|^2-4M\omega_2-4\omega_2^2\right)}{m_1^2}\ln\frac{m_\rho^2+\left(|p_t|-|q_t|\right)^2}{m_\rho^2+\left(|p_t|+|q_t|\right)^2}\\
&+12g_{DD\mathbb{V}}g_{D^*D^*\mathbb{V}}\left[m_\rho^2+2\left(|p_t|^2+|q_t|^2-2M\omega_2-2\omega_2^2\right)\right]\ln\frac{m_\rho^2+\left(|p_t|-|q_t|\right)^2}{m_\rho^2+\left(|p_t|+|q_t|\right)^2}\\
&+48g_\sigma^2m_1m_2\ln\frac{m_\sigma^2+(|p_t|-|q_t|)^2}{m_\sigma^2+(|p_t|+|q_t|)^2}
 +\frac{4g_{DD^*\mathbb{P}}^2|p_t||q_t|\left(\Lambda^2-m_\eta^2\right)\left(\Lambda^2-|p_t|^2+|q_t|^2\right)^2}{m_1^2\left[\Lambda^2+\left(|p_t|-|q_t|\right)^2\right]\left[\Lambda^2+\left(|p_t|+|q_t|\right)^2\right]}\\
&+\frac{12g_{DD^*\mathbb{P}}^2|p_t||q_t|\left(\Lambda^2-m_\pi^2\right)\left(\Lambda^2-|p_t|^2+|q_t|^2\right)^2}{m_1^2\left[\Lambda^2+\left(|p_t|-|q_t|\right)^2\right]\left[\Lambda^2+\left(|p_t|+|q_t|\right)^2\right]}
 +\frac{192g_\sigma^2m_1m_2|p_t||q_t|\left(\Lambda^2-m_\sigma^2\right)}{\left[\Lambda^2+\left(|p_t|-|q_t|\right)^2\right]\left[\Lambda^2+\left(|p_t|+|q_t|\right)^2\right]}\\
&-\frac{48g_{DD\mathbb{V}}g_{D^*D^*\mathbb{V}}|p_t||q_t|\left(\Lambda^2-m_\omega^2\right)\left(|p_t|^2-|q_t|^2\right)^2}{\left[\Lambda^2+\left(|p_t|-|q_t|\right)^2\right]\left[\Lambda^2+\left(|p_t|+|q_t|\right)^2\right]}
 -\frac{48g_{DD\mathbb{V}}g_{D^*D^*\mathbb{V}}\left(\Lambda^2-m_\rho^2\right)\left(|p_t|^2-|q_t|^2\right)^2}{\left[\Lambda^2+\left(|p_t|-|q_t|\right)^2\right]\left[\Lambda^2+\left(|p_t|+|q_t|\right)^2\right]}\\
&+\frac{48f_{DD^*\mathbb{V}}^2|p_t||q_t|\left(\Lambda^2-m_\omega^2\right)\left[\Lambda^4\left(M^2+8|p_t|^2\right)+2M^2\Lambda^2\left(|q_t|^2-3|p_t|^2\right)+M^2\left(|p_t|^2-|q_t|^2\right)^2\right]}{\left[\Lambda^2+\left(|p_t|-|q_t|\right)^2\right]\left[\Lambda^2+\left(|p_t|+|q_t|\right)^2\right]}\\
&+\frac{48f_{DD^*\mathbb{V}}^2|p_t||q_t|\left(\Lambda^2-m_\rho^2\right)\left[\Lambda^4\left(M^2+8|p_t|^2\right)+2M^2\Lambda^2\left(|q_t|^2-3|p_t|^2\right)+M^2\left(|p_t|^2-|q_t|^2\right)^2\right]}{\left[\Lambda^2+\left(|p_t|-|q_t|\right)^2\right]\left[\Lambda^2+\left(|p_t|+|q_t|\right)^2\right]}\\
&+24g_{DD\mathbb{V}}g_{D^*D^*\mathbb{V}}\left(|p_t|^2-|q_t|^2\right)^2\ln\frac{\Lambda^2+(|p_t|-|q_t|)^2}{\Lambda^2+(|p_t|+|q_t|)^2}
 +\frac{g_{DD^*\mathbb{P}}^2\left(\Lambda^2-2m_\eta^2+|p_t|^2-|q_t|^2\right)\left(\Lambda^2-|p_t|^2+|q_t|^2\right)}{|p_t|^2}\ln\frac{\Lambda^2+(|p_t|-|q_t|)^2}{\Lambda^2+(|p_t|+|q_t|)^2}\\
&+\frac{3g_{DD^*\mathbb{P}}^2\left(\Lambda^2-2m_\pi^2+|p_t|^2-|q_t|^2\right)\left(\Lambda^2-|p_t|^2+|q_t|^2\right)}{|p_t|^2}\ln\frac{\Lambda^2+(|p_t|-|q_t|)^2}{\Lambda^2+(|p_t|+|q_t|)^2}\\
&-\frac{12f_{DD^*\mathbb{V}}^2\left[-\Lambda^4\left(M^2+8|p_t|^2\right)+2\Lambda^2m_\omega^2\left(M^2+8|p_t|^2\right)+M^2m_\omega^2\left(2|q_t|^2-6|p_t|^2\right)+M^2\left(|p_t|^2-|q_t|^2\right)^2\right]}{|p_t|^2}\ln\frac{\Lambda^2+(|p_t|-|q_t|)^2}{\Lambda^2+(|p_t|+|q_t|)^2}\\
&-\frac{12f_{DD^*\mathbb{V}}^2\left[-\Lambda^4\left(M^2+8|p_t|^2\right)+2\Lambda^2m_\rho^2\left(M^2+8|p_t|^2\right)+M^2m_\rho^2\left(2|q_t|^2-6|p_t|^2\right)+M^2\left(|p_t|^2-|q_t|^2\right)^2\right]}{|p_t|^2}\ln\frac{\Lambda^2+(|p_t|-|q_t|)^2}{\Lambda^2+(|p_t|+|q_t|)^2}\\
&+\frac{g_{DD^*\mathbb{P}}^2\left(m_\eta^2-|p_t|^2+|q_t|^2\right)^2}{|p_t|^2}\ln\frac{m_\eta^2+\left(|p_t|-|q_t|\right)^2}{m_\eta^2+\left(|p_t|+|q_t|\right)^2}
 -12g_{DD\mathbb{V}}g_{D^*D^*\mathbb{V}}\left(|p_t|^2-|q_t|^2\right)^2\ln\frac{m_\omega^2+\left(|p_t|-|q_t|\right)^2}{m_\omega^2+\left(|p_t|+|q_t|\right)^2}\\
&+\frac{12f_{DD^*\mathbb{V}}^2\left[8|p_t|^2m_\omega^4+M^2m_\omega^4+M^2m_\omega^2\left(2|q_t|^2-6|p_t|^2\right)+M^2\left(|p_t|^2-|q_t|^2\right)^2\right]}{|p_t|^2}\ln\frac{m_\omega^2+\left(|p_t|-|q_t|\right)^2}{m_\omega^2+\left(|p_t|+|q_t|\right)^2}
 +\frac{3g_{DD^*\mathbb{P}}^2\left(m_\pi^2-|p_t|^2+|q_t|^2\right)^2}{|p_t|^2}\ln\frac{m_\pi^2+\left(|p_t|-|q_t|\right)^2}{m_\pi^2+\left(|p_t|+|q_t|\right)^2}\\
&-12g_{DD\mathbb{V}}g_{D^*D^*\mathbb{V}}\left(|p_t|^2-|q_t|^2\right)^2\ln\frac{m_\rho^2+\left(|p_t|-|q_t|\right)^2}{m_\rho^2+\left(|p_t|+|q_t|\right)^2}
 +\frac{12f_{DD^*\mathbb{V}}^2\left[8|p_t|^2m_\rho^4+M^2m_\rho^4+M^2m_\rho^2\left(2|q_t|^2-6|p_t|^2\right)+M^2\left(|p_t|^2-|q_t|^2\right)^2\right]}{|p_t|^2}\ln\frac{m_\rho^2+\left(|p_t|-|q_t|\right)^2}{m_\rho^2+\left(|p_t|+|q_t|\right)^2}\\
&+\frac{4g_{DD^*\mathbb{P}}^2|p_t||q_t|\left(\Lambda^2-m_\eta^2\right)\left(\Lambda^2-|p_t|^2+|q_t|^2\right)^2}{|p_t|^2\left[\Lambda^2+\left(|p_t|-|q_t|\right)^2\right]\left[\Lambda^2+\left(|p_t|+|q_t|\right)^2\right]}
 +\frac{12g_{DD^*\mathbb{P}}^2|p_t||q_t|\left(\Lambda^2-m_\pi^2\right)\left(\Lambda^2-|p_t|^2+|q_t|^2\right)^2}{|p_t|^2\left[\Lambda^2+\left(|p_t|-|q_t|\right)^2\right]\left[\Lambda^2+\left(|p_t|+|q_t|\right)^2\right]}\Bigg\}.
\end{split}
\end{equation}
\end{tiny}

\end{appendix}
\end{document}